\documentstyle[11pt,aaspp4]{article}
\begin{document}
\newcommand{\etal}{et al.}
\newcommand{\ie}{{i.e.}}
\newcommand{\eg}{{e.g.}}
\newcommand{\Msun}{\mbox{$M_\odot$}}
\newcommand{\hikpc}{\mbox{$h^{-1}$ kpc}}
\newcommand{\himpc}{\mbox{$h^{-1}$ Mpc}}
\newcommand{\Sub}[2]{\mbox{$#1_{\mbox{\scriptsize #2}}$}}
\newcommand{\Sup}[2]{\mbox{$#1^{\mbox{\scriptsize #2}}$}}
\newcommand{\beq}{\begin{equation}}
\newcommand{\eeq}{\end{equation}}
\newcommand{\beqa}{\begin{eqnarray}}
\newcommand{\eeqa}{\end{eqnarray}}
\newcommand{\lp}{\left(}
\newcommand{\rp}{\right)}

\title{Cosmological Simulations with Scale-Free Initial Conditions I:
Adiabatic Hydrodynamics}
\author{J. Michael Owen \altaffilmark{1}}
\affil{LLNL, L-16, P.O. Box 808, Livermore, CA 94551 \\
Email:  mikeowen@llnl.gov}
\altaffiltext{1}{Previous Address:  Ohio State University, Department of
Astronomy, Columbus, OH 43210}
\author{David H. Weinberg}
\affil{Ohio State University, Department of Astronomy, Columbus, OH 43210 \\
Email:  dhw@astronomy.ohio-state.edu}
\author{August E. Evrard \altaffilmark{2}}
\affil{University of Michigan, Physics Department, Ann Arbor, MI 48109 \\
Email: evrard@umich.edu}
\altaffiltext{2}{Also: Institut d'Astrophysique, 98bis Blvd. Arago, 75014,
Paris, France}
\author{Lars Hernquist}
\affil{University of California, Lick Observatory, Santa Cruz, CA 95064 \\
Email: lars@ucolick.org}
\author{Neal Katz}
\affil{University of Massachusetts, Department of Physics and Astronomy,
Amherst, MA  01003 \\ 
Email: nsk@kaka.phast.umass.edu}

\begin{abstract}
We analyze hierarchical structure formation based on scale-free initial
conditions in an Einstein-de Sitter universe, including a baryonic
component with $\Sub{\Omega}{bary} = 0.05$.  We present three independent,
smoothed particle hydrodynamics (SPH) simulations, performed at two
resolutions ($32^3$ and $64^3$ dark matter and baryonic particles), and
with two different SPH codes (TreeSPH and P3MSPH).  Each simulation is
based on identical initial conditions, which consist of Gaussian
distributed initial density fluctuations that have a power-spectrum $P(k)
\propto k^{-1}$.  The baryonic material is modeled as an ideal gas subject
only to shock heating and adiabatic heating and cooling; radiative cooling
and photoionization heating are not included.  The evolution is expected to
be self-similar in time, and under certain restrictions we identify the
expected scalings for many properties of the distribution of collapsed
objects in all three realizations.  The distributions of dark matter
masses, baryon masses, and mass and emission weighted temperatures scale
quite reliably.  However, the density estimates in the central regions of
these structures are determined by the degree of numerical resolution.  As
a result, mean gas densities and Bremsstrahlung luminosities obey the
expected scalings only when calculated within a limited dynamic range in
density contrast.  The temperatures and luminosities of the groups show
tight correlations with the baryon masses, which we find can be
well-represented by power-laws.  The Press-Schechter (PS) approximation
predicts the distribution of group dark matter and baryon masses fairly
well, though it tends to overestimate the baryon masses.  Combining the PS
mass distribution with the measured relations for $T(M)$ and $L(M)$
predicts the temperature and luminosity distributions fairly accurately,
though there are some discrepancies at high temperatures/luminosities.  In
general the three simulations agree well for the properties of resolved
groups, where a group is considered resolved if it contains more than 32
particles.
\end{abstract}

\section{Introduction}
The evolution of large-scale cosmological structure from scale-free initial
conditions has been well studied, and this idealized scenario provides many
useful insights into the general problem of gravitationally driven,
hierarchical structure formation.  So long as the background cosmology,
input physics, and initial conditions remain scale-free, such systems
should evolve self-similarly in time.  In reality the universe imposes
fixed physical scales (such as breaks in the initial power-spectrum of
density fluctuations, or gas cooling times), and therefore disrupts formal
self-similarity at some level.  However, for many interesting mass ranges,
such as those applicable to clusters of galaxies, the dominant physical
processes are essentially scale-free, and therefore the evolution of
structure on such scales may be approximately self-similar.  In this paper,
we use cosmological simulations with hydrodynamics to investigate
self-similar evolution of structure in a mixed baryon/dark matter universe.
Our goals are to investigate the physical properties of the structure that
evolves from scale-free initial conditions and to test the ability of
current cosmological hydrodynamics codes to reproduce analytically
predicted scalings.

Scale-free models are particularly useful in cosmology because in the
general case structure formation is analytically intractable, due to the
inherent nonlinearity of gravitational clustering.  If structure grows in a
self-similar manner, on the other hand, we can use scaling analysis to
predict how a given distribution of physical properties should evolve over
time, even though we cannot precisely predict what the quantitative values
of those properties should be. For example, while we may not know the
precise form of the mass distribution function of collapsed objects $f(M)$,
we can predict how this distribution should evolve with time, $f(M,t_1) \to
f(M,t_2)$.  Kaiser (1986) exploits this property in order to predict the
temporal evolution of the hot intracluster medium in galaxy clusters.
Note that this self-similarity of temporal evolution does not imply that
the structure present at any given time is spatially self-similar
(fractal), or that individual structures grow in the self-similar manner
described by Fillmore \& Goldreich (1984) and Bertschinger (1985ab).

Numerically modeling self-similar scenarios provides a particularly
powerful tool for cosmological investigations, since the numerical model
{\em can} provide detailed information about the state of the system at a
particular time, and then self-similarity can be used to scale this state
to any desired time.  Efstathiou \etal\ (1988) use collisionless N-body
simulations with scale-free initial conditions to study structure formation
for a variety of input power-spectra, testing their results for the
expected self-similar scalings.  Efstathiou \etal\ (1988) also compare
their results to those of approximate analytic models such as those of
Davis \& Peebles (1977) and Press \& Schechter (1974).

Here we present a set of hydrodynamic simulations based on scale-free
initial conditions, in the hope that we will be able to identify
self-similar scaling in the properties of both the gas and the
collisionless dark matter.  In the absence of radiative cooling, adding
hydrodynamical processes to the purely gravitational systems studied
previously does not introduce any new physical scales, and therefore the
properties of the baryons should scale self-similarly as well as those of
the dark matter.  This is not to say that the two species should evolve
identically to one another.  The physics governing the details of collapse
in the two species, with the baryons subject to added processes such as
shock heating and pressure support, are quite different.  Self-similarity
does imply that given the detailed differences in the arrangement of the
dark matter and baryons in a spectrum of structures at a given time,
corresponding structures at any other time should be arranged in the same
manner.  Basic observational quantities such as the cluster X-ray
luminosity function depend upon how the baryons are distributed in
collapsed, cluster-scale objects.  Additionally, there is reason to suspect
that numerical effects may be more (or at least differently) evident in the
baryons than in the dark matter.  Several independent investigations
(Anninos \& Norman 1996; Weinberg, Hernquist, \& Katz 1997; Owen \&
Villumsen 1997; to name merely a few) have found evidence for complicated
numerical artifacts in the properties of baryons in collapsed structures --
particularly in the innermost, high-density regions.  However, because they
are connected to numerical parameters held fixed during the evolution of
the system (\eg, the number of particles), such numerical effects should
not scale self-similarly, and we should be able to use departures from
self-similar scaling of the physical properties of the system to identify
regimes in which the numerical effects dominate.  The self-similar scaling
identified in the collisionless experiments of Efstathiou \etal\ (1988) is
one of the key pieces of evidence for the reliability of N-body methods as
a tool for investigating cosmological structure formation.  This paper
attempts to apply similar tests to hydrodynamical simulations, performed
with the TreeSPH (Hernquist \& Katz 1989, hereafter HK89; Katz, Weinberg,
\& Hernquist 1996, hereafter KWH96) and P3MSPH (Evrard 1988) simulation
codes.

Even in the adiabatic limit, there are some rather subtle physical scales
set by the physics of the gas, which could affect measurable quantities of
interest.  For instance, the TreeSPH code calculates the
temperature-dependent mean molecular weight of the gas throughout its
evolution.  The temperature-dependent ionization state of the gas does not
directly affect the gas dynamics, as it is the specific thermal energy, not
the temperature, that determines the gas pressure.  However, changing the
mean molecular weight {\em does} influence the mapping of the specific
thermal energy to temperature (by a factor $\sim 2$ for temperatures in the
range $T \in [10^4$K$, 10^5$K$]$), and also such observationally
interesting quantities as the Bremsstrahlung luminosity.  Throughout this
paper, whenever the mean molecular weight $\mu$ explicitly enters into an
equation we will assume $\mu = 0.6$, appropriate for a fully ionized,
primordial H/He gas.  Though the TreeSPH code does use the variable $\mu$
in order to determine the gas temperature, this is a small effect in the
temperature range relevant to our simulations, and it does not
significantly influence our results.

Radiative cooling is an important process in the formation and evolution of
galaxies, but it plays a less significant role on the scale of galaxy
clusters.  Therefore, the results of this paper are most directly relevant
to cluster studies, which are the focus of many recent hydrodynamical
investigations, such as Bryan \etal\ (1994), Cen \etal\ (1995), Evrard,
Metzler, \& Navarro (1996), and Pinkney \etal\ (1996).  While imposing a
realistic radiative cooling law violates self-similarity, it is possible to
construct artificial cooling laws that do not.  In a subsequent
investigation we will present the results of self-similar studies including
radiative cooling, in order to more directly address the issue of galaxy
formation.

This paper is organized as follows.  In \S \ref{Sims.sec} we discuss how
the simulations are performed and analyzed.  We consider experiments
performed with two different cosmological hydrodynamical codes (TreeSPH and
P3MSPH) and at two different resolutions ($64^3$ and $32^3$ baryon and dark
matter particles).  In \S \ref{SelfSim.sec}, the core of the paper, we
summarize the expected scalings and test the simulations to see how well
various physical properties of the collapsed objects follow these
expectations.  Finally, in \S \ref{DirectComp.sec} we perform direct
comparisons of the different experiments, in order to investigate how well
the different numerical models reproduce the same physical results.
Section \ref{Summary.sec} summarizes our results.

\section{The Simulations}
\label{Sims.sec}
We analyze three 3-D hydrodynamical simulations based on the same initial
conditions.  The background cosmology is a flat, Einstein-de Sitter
universe with $\Sub{\Omega}{bary}=0.05$ and $\Sub{\Omega}{dm}=0.95$.  The
initial density perturbations are Gaussian with a power-law power-spectrum
$P(k) \propto k^{-1}$.  The initial amplitude is chosen so that the
linearly predicted mass fluctuations will have an rms value of unity within
a top-hat window of radius $\Sub{R}{th} \approx 0.2 L_0$ at the end of the
simulation, where $L_0$ is the simulation box size.  The initial density
fluctuations are similar to those used in Katz, Hernquist, and Weinberg
(1992) and KWH96, except that here we use a pure $n=-1$ power-spectrum for
the density fluctuations rather than that of a Cold Dark Matter (CDM)
model.  Throughout this paper we parameterize the evolution in terms of the
expansion factor $a$, defined so that the final output expansion is $a_f
\equiv 1$.  In principle, because of the scale-free nature of these
experiments, there are many possible choices for identifying the simulation
box size with a physical scale in the real universe.  For instance, if we
identify the $a = a_f \equiv 1$ output with the observed universe at $z=0$
and adopt the normalization condition $\sigma_8 = 0.5$ (appropriate for the
rms mass fluctuation in spheres of radius 8 \himpc\ with $\Omega = 1$; see
White, Efstathiou, \& Frenk 1993), then the implied comoving box size is
$L_0 \approx 20 \himpc$, and earlier output times can be identified with
higher redshifts, $z = a^{-1} - 1$.  With this scaling to physical units,
the computational box size represents a rather small cosmological volume,
and the ``clusters'' that form are more representative of galaxy groups
rather than rich clusters such as Coma.  Alternatively, one can identify
any expansion factor $a$ with the $z=0$ universe, in which case the implied
box size (for $\sigma_8 = 0.5$) is $L_0 = 20 ~a^{-1} ~\himpc$, decreasing
steadily as the nonlinear scale $0.2 L_0$ becomes a larger fraction of the
box size.

Within this framework we perform a low-resolution experiment (where the
number of baryonic and dark matter particles is
$\Sub{N}{bary}=\Sub{N}{dm}=32^3$) with TreeSPH (HK89; KWH96), and two
high-resolution experiments ($\Sub{N}{bary}=\Sub{N}{dm}=64^3$) using
TreeSPH and P3MSPH (Evrard 1988).  We examine each simulation at expansions
spaced logarithmically with an interval of $\Delta \log a=0.2$.  At each
expansion we identify groups of particles using the ``friends-of-friends''
algorithm (see, \eg, Barnes \etal\ 1985; the specific implementation used
here can be found at ``http://www-hpcc.astro.washington.edu/tools/FOF/'').
We compute global average properties for each group (such as the dark
matter and baryon mass, the mass and emission weighted temperatures,
luminosity, and so forth) and check for self-similar behavior in the
distribution of properties among these groups.  Though there are more
sophisticated group finding algorithms available, friends-of-friends
provides a simple, flexible, and unambiguous definition of a group using an
algorithm that maintains the conditions necessary for self-similar scaling
(because it does not introduce a fixed physical scale).  For a given
linking parameter $l = b ~\Delta x_p$ (where $\Delta x_p$ is the initial
interparticle spacing), friends-of-friends selects groups, on average,
within an overdensity contour of roughly $\delta \rho/\bar{\rho} \approx
2~b^{-3}$.  Throughout this paper we use a linking parameter $l = 0.2
~\Delta x_p$, so that we are selecting objects roughly of overdensity
$\delta \rho/\bar{\rho} \approx 250$.

In order to understand the regimes we can probe, we must first identify the
mass resolution limits of our experiments.  Since SPH is a Lagrangian
technique, the lower limit on the hydrodynamic interactions is best
expressed as a mass limit, set by a multiple of the particle mass.  In each
simulation the SPH smoothing scales are evolved so that each particle
samples roughly 32 of its neighbors, which provides a reasonable
lower-limit on the SPH mass resolution.  The gravitational force resolution
is effectively determined by a multiple of the gravitational softening
length, so unfortunately this is not Lagrangian in the same way as the
hydrodynamic resolution.  However, in each run the gravitational softening
length is $\Sub{L}{soft} = \Sub{L}{box}/1111$, and on the scale of the
objects we will be examining we may consider the gravitational resolution
to be unrestrictive.  The minimum resolved mass is therefore set by the 32
particle limit.

\begin{figure}[htbp]
\epsscale{0.5}
\plotone{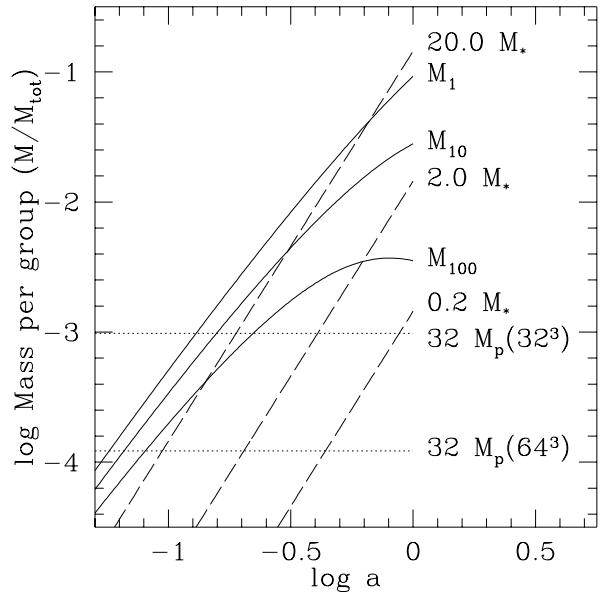}
\epsscale{1}
\caption{Mass resolution limits as a function of expansion factor.  The
dotted lines show 32 times the particle mass for the different simulation
resolutions ($N$ = $32^3$ and $64^3$ particles).  The dashed lines show
various multiples of $M_*$, the characteristic mass for an object with a
linearly predicted overdensity of $(\delta \rho/\bar{\rho}\Sub{)}{lin} =
1.68$.  The solid lines show the group masses above which the
Press-Schechter formalism predicts an average of 1 ($M_1$), 10 ($M_{10}$),
and 100 ($M_{100}$) groups in our simulation volume.  Mass units are scaled
to the total mass within the simulation volume.}
\label{MassRes.fig}
\end{figure}
The upper limit on the mass range we are sensitive to is set by our box
size.  The larger a group is, the more statistically rare it is, and the
less likely we are to find examples of such structures in any finite,
randomly realized volume.  We can therefore only expect to find objects up
to a certain size at any given expansion within our finite simulation
volume.  The Press-Schechter mass function (Press \& Schechter 1974)
provides a rough estimate of this limit.  In Figure \ref{MassRes.fig} we
plot the lower and upper mass limits over a range of expansions.  We also
show the evolution of various multiples of a ``characteristic'' mass $M_*$,
where $M_*$ is defined as the mass of an object with a linearly predicted
overdensity of $\delta M/M = 1.68$.  This is chosen as an appropriate
comparison mass for our friends-of-friends identified groups, as is
discussed in more detail in \S \ref{ExpScale.sec}.  Based on Figure
\ref{MassRes.fig} we can reasonably expect the mass range $[0.2 ~M_*, 2.0
~M_*]$ to be accessible for expansions $\log a \in [-0.4, 0]$, though the
$32^3$ simulation may be dicey on the low end at early expansions.  Note
that this only represents about a factor of 2.5 in expansion, so we are
fairly restricted in the range of expansion factor that is suitable for
tests of self-similar scaling.  Future experiments with more particles
could certainly improve on this range, but the particle numbers used here
are fairly representative of the SPH experiments published to date.

\section{Self-similar evolution}
\label{SelfSim.sec}
\subsection{Expected scalings}
\label{ExpScale.sec}
Before we discuss in detail the results of the simulations, it is useful to
summarize the expected scalings.  Kaiser (1986) provides a good overview of
this topic, so we only briefly cover the salient points here.  Under our
scale-free assumption for the input physics and the background cosmology
(barring small effects such as the temperature dependence of $\mu$ noted
above), the only available physical scale is set by the amplitude of
density fluctuations, \eg, the length-scale on which the rms density
fluctuations have $\delta \rho/\bar{\rho} \approx 1$ at a given expansion.
This nonlinear scale can be equivalently expressed in terms of a length
\Sub{R}{nl} or a mass $\Sub{M}{nl} = 4/3 \pi \Sub{R}{nl}^3 \bar{\rho}$.  So
long as the system obeys temporal self-similarity, any dimensionless
statistic set by the density field must be a function of this nonlinear
scale only ($M/\Sub{M}{nl}$ or $R/\Sub{R}{nl}$). We test for self-similar
scaling in the simulations by following the evolution of distributions of
dimensionless variables such as $M/M_*$ or $T/T_*$, as functions of the
expansion factor $a$, where the variables subscripted with $*$ are the
``characteristic'' quantities, which evolve with $a$ according to the
analytically predicted scaling.

The evolution of these characteristic quantities can be parameterized in
terms of the power-law index of the density perturbations $n$, where $P(k)
\propto k^n$.  The characteristic length and mass scales follow the
corresponding nonlinear scales, given by
\beqa
  R_* &\propto& \Sub{R}{nl} \propto a^{(5 + n)/(3 + n)}, \\
  M_* &\propto& \Sub{M}{nl} \propto a^{6/(3 + n)}.
\eeqa
The characteristic density evolves in proportion to the background density,
\beq
  \rho_* \propto \frac{M_*}{R_*^3} \propto \bar{\rho} \propto a^{-3},
\eeq
while the characteristic temperature and Bremsstrahlung luminosity are
defined through combinations of these parameters,
\beqa
  T_* &\propto& \frac{M_*}{R_*} \propto a^{(1 - n)/(3 + n)}, \\
  L_* &\propto& M_* \rho_* T_*^{1/2} \propto a^{-(5 + 7n)/(6 + 2n)}.
\eeqa
Our initial conditions have a spectral index $n=-1$, so these scalings
become $R_* \propto a^2$, $M_* \propto a^3$, $\rho_* \propto a^{-3}$, $T_*
\propto a^1$, and $L_* \propto a^{1/2}$.  Note that these scalings are
expressed in proper coordinates.

Given the relative scalings, we must now choose reasonable normalizations
for our characteristic parameters.  Since the friends-of-friends algorithm
identifies particle groups with overdensity $\delta \rho/\bar{\rho} \approx
250$, we will try to select parameters characteristic of such objects.  The
spherical top-hat model for the collapse of an isolated perturbation
predicts that an overdense region will collapse when it reaches a linearly
predicted overdensity of $(\delta \rho/\bar{\rho}\Sub{)}{lin} = 1.68$ --
its actual density contrast at this point is of order 200, roughly that of
our fiducial objects identified with friends-of-friends.  Our initial
density perturbations are normalized so that the linearly predicted rms
overdensity in a top-hat of radius 0.2 \Sub{L}{box} at the final expansion
is $\sigma(a_f) \approx 1$, and for a $P(k) \propto k^{-1}$ power-spectrum,
$\sigma^2 \propto R^{-(n + 3)} = R^{-2}$.  Using this information, we can
define characteristic quantities that are roughly appropriate for objects
of linearly predicted overdensity $(\delta \rho/\bar{\rho}\Sub{)}{lin} =
1.68$ in species $X$ as
\beqa
  R_* &\equiv& a^{2} ~0.120 \ \Sup{L_f}{box}, \\
  M_* &\equiv& a^{3} ~7.25 \times 10^{-3} \Omega_X \ \Sup{M}{box}, \\
  \rho_* &\equiv& a^{-3} ~1.00 ~\Omega_X \ \Sup{M}{box} (\Sup{L_f}{box})^{-3},\\
  \label{T1.eq}
  T_* &\equiv& \frac{\mu m_p}{2 k_B} \frac{G M_*}{R_*} = a ~6.04 \times 10^{-2}
      ~\frac{\Sup{M}{box}}{\Sup{L_f}{box}}~\frac{\mu m_p G}{2 k_B}.
\eeqa
For convenience, we define the characteristic Bremsstrahlung luminosity
directly in terms of $M_*$, $\rho_*$, and $T_*$,
\beq
  L_* \equiv M_* \rho_* T_*^{1/2},
\eeq
without the physical constants that convert this to luminosity units.
However, we can express $L_*$ in c.g.s.\ units using the Bremsstrahlung
volume emissivity for a primordial composition plasma, $\epsilon
= 1.42 \times 10^{-27} ~T^{1/2} ~(\Sub{n}{H$^+$} + \Sub{n}{He$^+$} + 4
\Sub{n}{He$^{++}$}) n_e$ erg sec$^{-1}$ cm$^{-3}$ (Black 1981; we have set
the Gaunt factor $\Sub{g}{ff} = 1$).  For a helium mass fraction $Y =
0.24$, we obtain
\beq
  L_* = \frac{4}{3} \pi R_*^3 \epsilon_* = a^{1/2} 
        ~4.24 \times 10^7 ~\Sub{\Omega^2}{bary} 
        ~\lp \frac{\Sup{M}{box}}{\Msun} \rp^{5/2}
	\lp \frac{\Sup{L_f}{box}}{\himpc} \rp^{-7/2}
        h ~\mbox{ergs sec$^{-1}$}.
\eeq

If we scale to a physical box size $L_0 = 20 \himpc$ as outlined above,
then these quantities become $R_* = 2.40 ~a^2 ~\himpc$, $M_* = 1.61 \times
10^{13} ~a^{3} ~\Omega_X h^{-1}$ \Msun, $\rho_* = 2.78 \times 10^{11}
~a^{-3} ~\Omega_X h^2$ \Msun\ Mpc$^{-3}$, $T_* = 1.05 \times 10^6 ~a$ K,
and $L_* = 6.90 \times 10^{38} ~a^{1/2} ~h$ ergs sec$^{-1}$.

\subsection{Global scaling in the simulations}
\begin{figure}[htbp]
\plotone{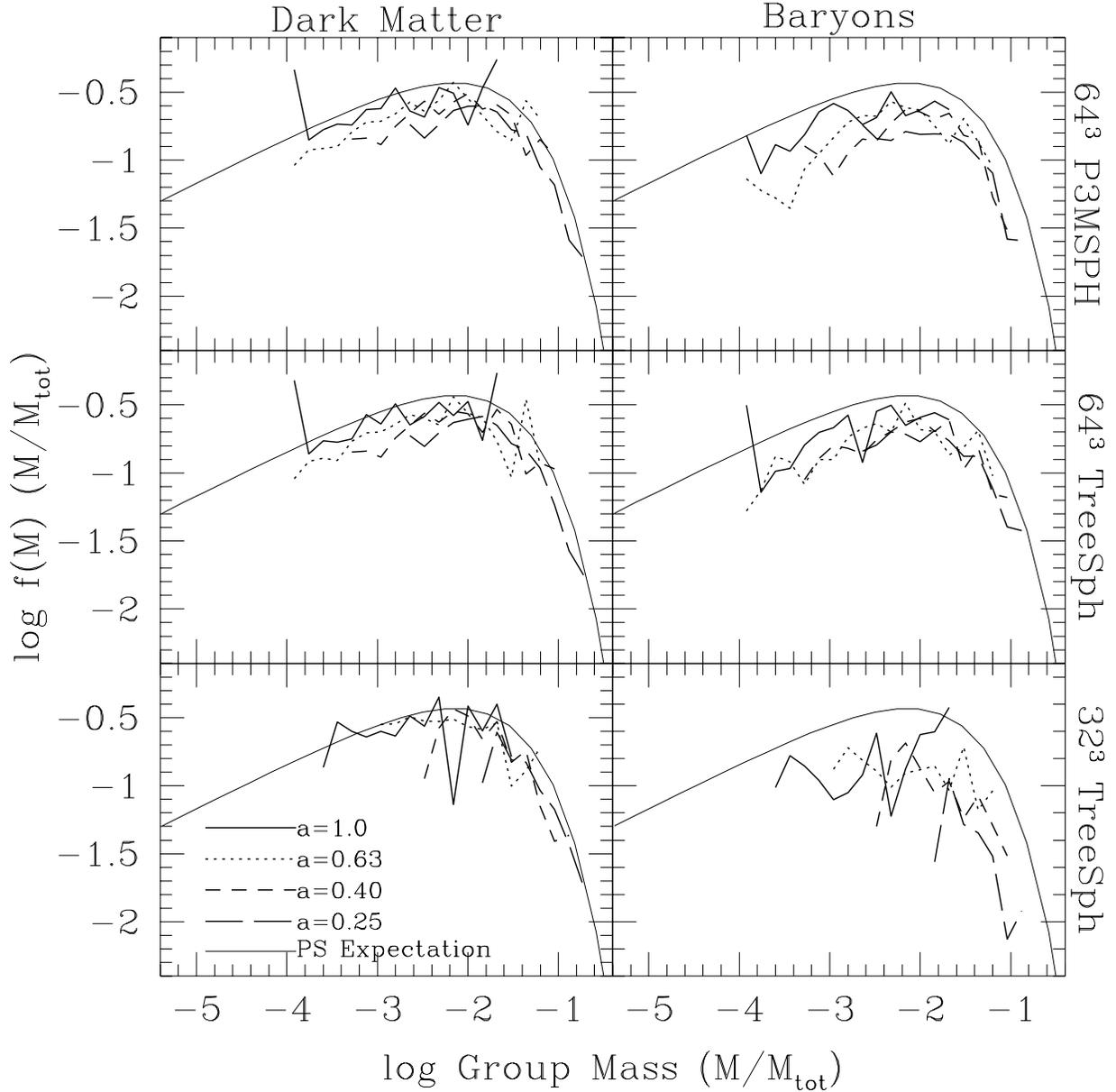}
\caption{The differential mass distribution function of group masses
$f(M)$ (baryons and dark matter plotted separately) for expansions $a$ =
0.25, 0.4, 0.63, and 1.0.  All quantities are scaled self-similarly to $a =
a_f \equiv 1$.  Each expansion is represented by a different line type,
with the Press-Schechter (PS) prediction plotted as the thin solid line.
The masses are scaled so that the total mass in each species is 1.}
\label{fM.fig}
\end{figure}
Perhaps the most obvious statistic we can measure is the differential
distribution of the mass in groups, $f(M)$.  Figure \ref{fM.fig} shows
$f(M)$ for each of these simulations, plotting the dark matter and baryons
separately.  Note that we compute the amount of mass contained in groups in
each mass range, not the number of groups.  We measure $f(M)$ at four
different expansion factors and scale the results self-similarly to $a=a_f
\equiv 1$ for comparison.  Thus, if the simulations perfectly obeyed the
analytically predicted, self-similar scalings, the plotted mass functions
would lie on top of one another within the measured mass ranges.  There are
no free parameters in this scaling, and the measured distributions $f(M)$
at different expansions appear to link up quite nicely.  Therefore, it
appears that the mass is following the expected self-similar behavior
reasonably well, though somewhat better in the dark matter than the
baryons.  As we might anticipate from Figure \ref{MassRes.fig}, it is clear
that the simulations probe different regions of the overall mass function
at different expansions.  At early times only the highest mass objects have
collapsed, and therefore we tend to see only the high end of $f(M)$
represented.  As the simulations evolve we progressively resolve smaller
and smaller objects in the overall mass range, until by the end we begin to
lose the high mass end due to the limited size of the computational box.

The thin solid lines in Figure \ref{fM.fig} show the mass function
predicted by the Press-Schechter (PS) formalism.  Note that this prediction
has no free parameters, so these curves are not fits to the numerical data.
In general the mass distribution follows the shape of the PS prediction
quite well, though the baryon mass in groups tends to be a bit low compared
with both the PS prediction and the dark matter results.  This depression
in the baryon mass in groups across the entire measured mass range suggests
that the baryon fraction in collapsed objects slightly underrepresents the
universal baryon to dark matter mixture, at least within a density contrast
of $\delta \rho/\bar{\rho} \approx 250$.  This result supports similar
findings by previous studies (Evrard 1990; Kang \etal\ 1994a), suggesting
that the baryon fraction in collapsed objects is depleted relative to the
universal mixture, at least in the absence of cooling.  Still, these
results must be treated with some caution, since there are indications that
numerical effects due to finite resolution can drive the baryon to dark
matter ratio in this direction (Owen \& Villumsen 1997).

\begin{figure}[htbp]
\epsscale{0.8}
\plotone{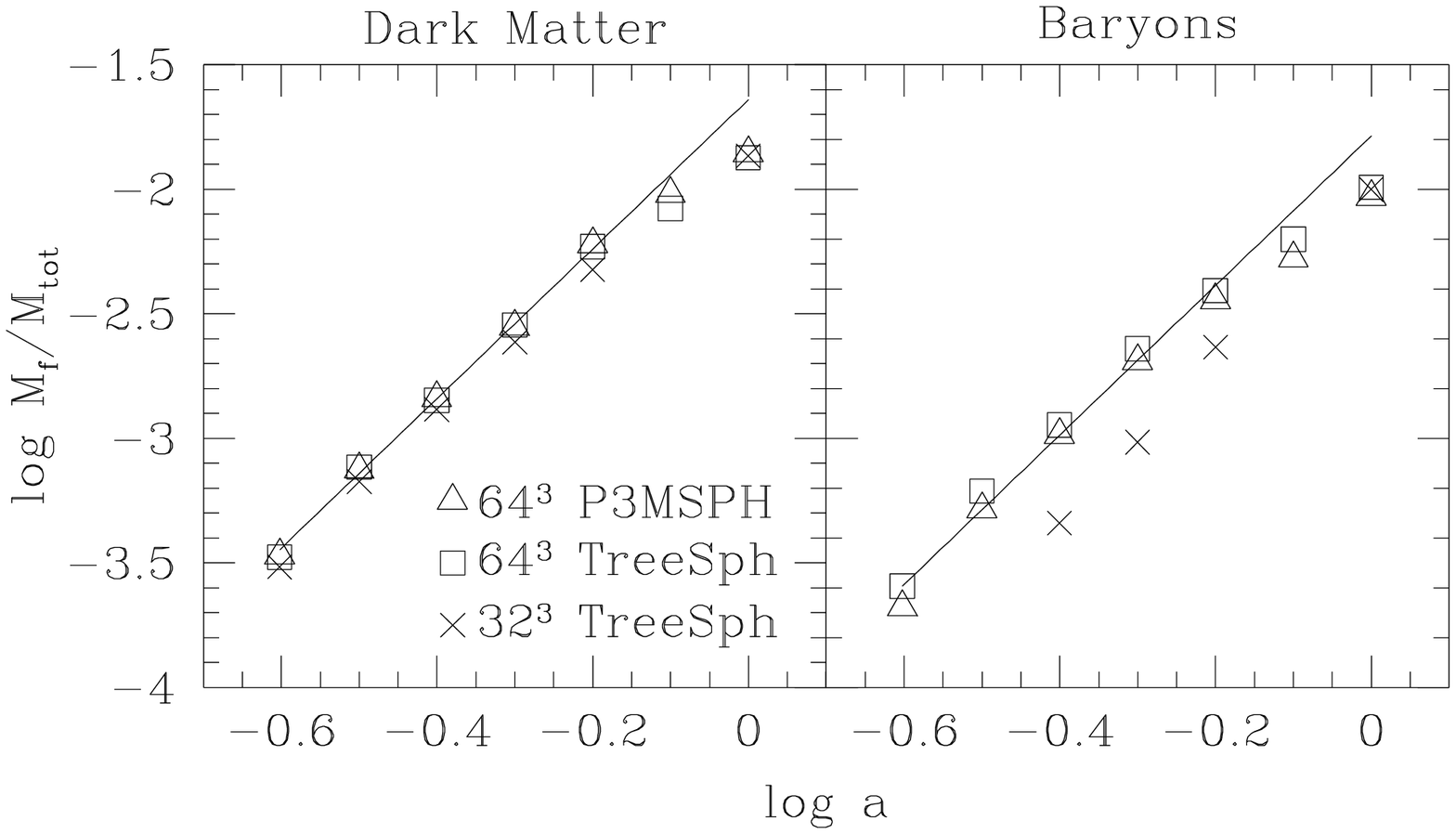}
\epsscale{1}
\caption{Evolution of a fiducial group mass $M_f(a)$, defined so that at each
expansion $90\%$ of the total mass in species $X$ (baryons and dark matter
plotted separately) is contained in groups of mass $M_X \le M_f$.  The
points depict the results of each simulation at specific expansions, and
the lines show the predicted power-law $M_f \propto a^3$ normalized to the
high-resolution experiments for $\log a \in [-0.6, -0.2]$.}
\label{Mscale.fig}
\end{figure}
In order to demonstrate more directly the self-similar scaling of the group
masses, Figure \ref{Mscale.fig} measures the fiducial mass $M_f$ defined so
that $90\%$ of the total mass is contained in groups of mass $M \le M_f$.
Isolated particles are counted as ``groups'' of mass $\Sub{M}{particle}$
for this purpose.  The solid lines show the self-similar scaling solution
$M_f \propto a^3$ normalized to the high-resolution experiments ($64^3$
TreeSPH and P3MSPH) in the expansion range $\log a \in [-0.6, -0.2]$.  This
normalization is determined by first scaling each measurement to a fiducial
time (assuming self-similarity), then taking the average of all of these
scaled measurements.  Each point plotted is determined by summing the
cumulative mass function, locating the two mass points that bracket our
desired mass fraction, and linearly interpolating (in log space) between
these points.  The observant reader will notice in this and subsequent
figures that the $\log a = -0.1$ output is missing from the $32^3$ TreeSPH
measurements.  That timeslice was inadvertently overwritten and lost in the
process of transferring and analyzing the data.

When considering this figure (as well as the following similar figures), it
is worthwhile to keep the mass resolution limits of these simulations in
mind.  Our measured value for $M_f$ is only valid so long as the
simulations can sample $f(M)$ in a statistically meaningful way near the
chosen mass fraction.  In a hierarchical structure formation scenario, all
of the mass may be bound in structures down to arbitrarily small mass
scales (Bond \etal\ 1991).  However, any simulation with finite mass
resolution can represent bound objects only down to its resolution limit --
the mass which should be bound in structures below this scale is simply not
represented in the simulations.  Likewise, the finite simulation volume
implies that we can statistically represent structures only up to a certain
mass limit.  As the simulations evolve we begin to lose the upper end of
the mass function because the power saturates in the longest wavelength
modes of the simulation.  We have chosen a fairly large mass fraction
($90\%$) in order to sample the distribution back to early expansions, when
only the upper end of the group mass distribution is accessible.  However,
as the evolution of the system progress and larger and larger scales become
nonlinear, we lose the high mass end of the distribution, which is why the
final few points at large expansions in Figure \ref{Mscale.fig} fall below
the extrapolation from earlier epochs.

Leaving aside these last two points, the group masses scale quite well in
the range of expansions $\log a \in [-0.6, -0.2]$.  The dark matter group
masses seem to obey self-similar scaling somewhat better than the baryons,
but the differences are minor.  The dark matter and baryon measurements in
Figure \ref{Mscale.fig} are normalized separately to the total amount of
mass in each species, so the fact that the baryon points at each epoch tend
to be slightly lower than the corresponding dark matter measurements
supports the trend noted in the $f(M)$ distributions in Figure \ref{fM.fig}
-- the fraction of baryons in collapsed structures is slightly depressed
compared with the collapsed fraction of dark matter, by roughly $10\%$ at
$a=a_f$.  Note also that while both of the high-resolution experiments seem
to agree on the masses quite well, there is a trend for the low-resolution
$32^3$ TreeSPH simulation to find systematically lower baryon masses.  This
suggests there may be resolution artifacts which tend to suppress the
baryon masses of collapsed structures, even though the dark matter masses
appear to be unaffected.

The baryon and dark matter groups plotted in Figures \ref{fM.fig} and
\ref{Mscale.fig} are identified by running the friends-of-friends algorithm
on each species separately, so we cannot examine the relative mixtures of
baryons and dark matter on a group-by-group basis.  In order to examine the
influence of resolution on baryon-to-dark matter ratios more directly, we
also applied the friends-of-friends algorithm to the full complement of
baryon and dark matter particles (recall that there are equal numbers of
particles of each species, with the dark matter particles being more
massive).  This procedure yields a set of baryon/dark matter groups, and we
can examine the baryon fraction as a function of the total object mass.  We
find that resolution does influence the baryon to dark matter ratio in
moderately resolved objects, in the sense that baryons are systematically
underrepresented in low mass, poorly resolved structures.  However, we also
find that for groups with more than $\sim 150$ SPH particles the baryon to
dark matter ratio plateaus to a constant value, roughly
$\Sub{n}{bary}/\Sub{n}{dm} \in [0.8, 0.9]$ in the high-resolution
experiments.  It thus appears that the typical baryon fraction in
virialized systems, at the overdensity level $\delta \rho/\bar{\rho} \sim
250$, is about $85\%$.  The scaling of the mass functions, consistency of
the baryon to dark matter ratio for well resolved groups, and agreement
between the two independent codes all suggest that this result is robust,
though confirmation with higher resolution experiments would be desirable.

\begin{figure}[htbp]
\epsscale{0.8}
\plotone{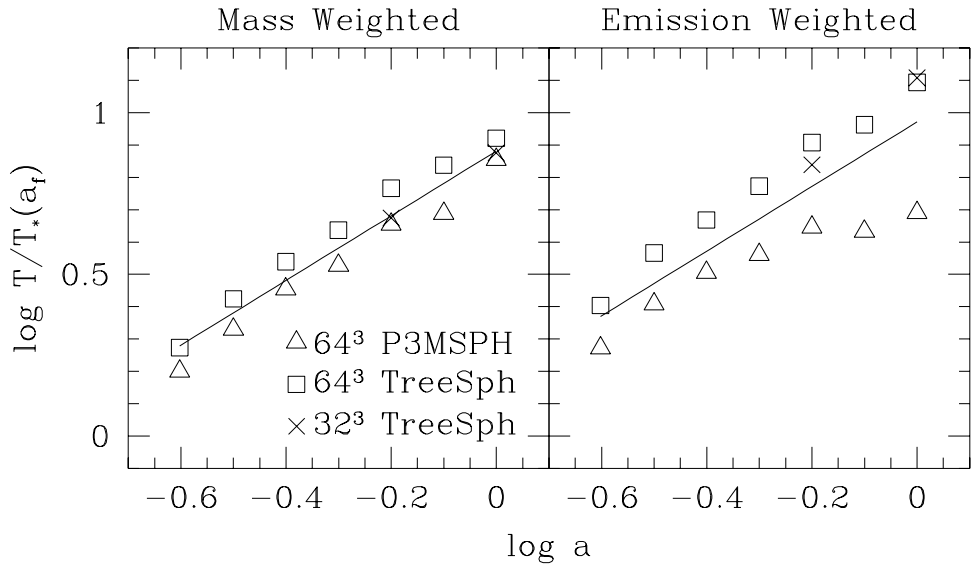}
\epsscale{1}
\caption{Evolution of the fiducial group temperature $T_f(a)$, defined so
that at each expansion $90\%$ of the total mass is contained in groups with
temperature $T \le T_f$.  The left panel uses the mass weighted
temperatures for each object, and the right the emission weighted.  The
solid line shows the expected self-similar evolution, normalized to the
average of the two high-resolution experiments for $\log a \in [-0.6,
-0.2]$, calculated similarly to those shown in Figure \ref{Mscale.fig}.
Temperatures are scaled to the characteristic temperature $T_*(a_f)$ at $a
= a_f \equiv 1$ (Eq. \ref{T1.eq}).}
\label{Tscale.fig}
\end{figure}
In Figure \ref{Tscale.fig} we plot the evolution of a fiducial group
temperature $T_f(a)$ defined in similar fashion to the fiducial mass used in
Figure \ref{Mscale.fig}.  The left panel shows the mass weighted
temperature, and the right shows the emission weighted temperature.  The
mass weighted temperature for a given group $i$ is simply the average
temperature over each particle $j$ that is a member of that group: $T_i =
N_i^{-1} \sum_j T_j$.  The emission weighted temperature is defined
assuming Bremsstrahlung radiation, for which the bolometric, volume
emissivity goes as $\epsilon \propto \rho^2 T^{1/2}$.  The emission
weighted temperature associated with group $i$ is then
\beq 
  \label{Tem.eq}
  T_i = \frac{\int \epsilon T ~dV}{\int \epsilon ~dV}
      = \frac{\int \rho^2 T^{3/2} ~dV}{\int \rho^2 T^{1/2} ~dV}
      = \frac{\sum_j m_j \rho_j T_j^{3/2}}{\sum_j m_j \rho_j T_j^{1/2}},
\eeq
again represented as a sum over all particles $j$ that are members of group
$i$.  We sort these individual group temperatures in ascending order, and
define the fiducial temperature $T_f(a)$ such that $90\%$ of the baryon mass
is contained in objects with $T_i \le T_f$.  As with the mass scaling shown
Figure \ref{Mscale.fig}, we count the mass contained in unresolved groups
as having temperatures less than the lowest measured group temperature.
The solid lines show the self-similar solution normalized to the
high-resolution results for $\log a \in [-0.6, -0.2]$.  Temperatures are
scaled to the characteristic temperature $T_*$ (eq.\ \ref{T1.eq}) defined
at $a = a_f \equiv 1$; if we scaled instead to $T_*(a)$ (the characteristic
temperature at each expansion), then the predicted evolution tracks would
be horizontal lines.

Considering first the mass weighted temperature evolution in the left
panel, we can see that both of the high-resolution experiments scale
equivalently, though the P3MSPH simulation tends to find slightly cooler
temperatures than TreeSPH.  The low-resolution TreeSPH experiment also
shows the expected scaling between $\log a = -0.2$ and $\log a = 0$.
During this final period the low-resolution temperatures are consistent
with the high-resolution measurements.  At earlier times, however, the
fiducial temperatures measured for the $32^3$ TreeSPH experiment are off
the scale of this figure, well below the plotted range.  The failure of the
low-resolution TreeSPH experiment at early expansions is not surprising
once we realize that with $N=32^3$ particles a typical $M_*$ group will not
be represented by more than 32 particles until expansion $\log a \gtrsim
-0.3$ (see Figure \ref{MassRes.fig}).  The evolution of the emission
weighted temperature (in the right panel) is similar to the mass weighted
behavior for TreeSPH, but not for P3MSPH.  Both of the high-resolution
experiments scale well in the emission weighted temperatures for $\log a
\lesssim -0.2$.  At later times, the high-resolution TreeSPH experiment
continues to follow the expected temperature scaling, but the P3MSPH
temperatures begin to fall off.  Additionally, we can see that the
temperature difference between the TreeSPH and P3MSPH simulations is larger
in the emission weighted temperature than in the mass weighted temperature.

In general, both the mass and emission weighted temperatures scale well.
However, the physical processes affecting these two quantities differ
slightly, so they are interesting to compare.  The mass weighted
temperature provides a fairly direct measure of the total energy converted
from kinetic form to thermal form through shocks in accreting structures.
The fact that the mass weighted temperatures scale quite effectively
indicates that the dissipation of energy through shocks is being modeled
well over a wide range of mass and length scales.  This is a nontrivial
test, as shocks are an extremely important process in determining the final
properties of the gas contained in collapsed objects.

The emission weighted temperature is weighted toward the highest density
regions in each collapsed object -- this follows because the volume
emissivity goes as $\epsilon \propto \rho^2$.  The highest density regions
in the collapsed objects are in their cores, so the emission weighted
temperature of each group is weighted toward the core temperature.  The
fact that the mass weighted temperatures agree closely between the P3MSPH
and TreeSPH experiments indicates that the differences noted in the total
emission weighted temperatures are largely due to differences in the cores
of these objects.  The group cores in the P3MSPH simulation are
characteristically cooler than those in the TreeSPH case.  Additionally,
the deviant scaling of the emission weighted temperatures in the P3MSPH run
at late times is likely due to the increasing importance of core
contributions to this measure.  The growing dynamic range in density
progressively resolves cores (\ie, regions above very high density contrast
$\delta \rho/\bar{\rho} \sim 10^4$) within smaller objects, leading to a
continuously increasing mass fraction in cores over time.  The cores in
the P3MSPH simulation exhibit temperature inversions, while those in the
TreeSPH simulation do not.  The effect of this structural difference
becomes apparent only at late times, when a sufficient mass fraction in the
cool cores has been resolved.  Note that the core mass fraction is quite
small even at the final epoch, as constrained by the rather good scaling of
mass weighted temperature exhibited within both codes.  We will return to
consider these points in more detail below.  Despite these subtle effects,
the mass weighted group temperatures scale quite effectively, indicating
that the simulations follow the important process of shock heating well.

\begin{figure}[htbp]
\plotone{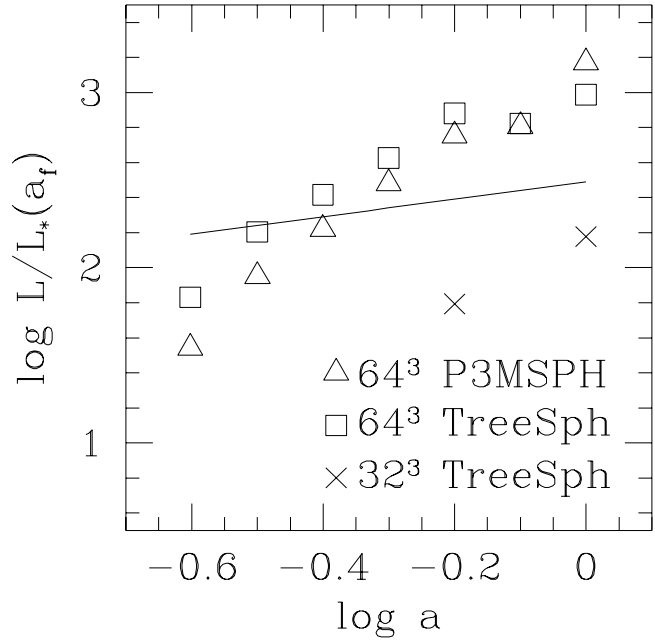}
\caption{Evolution of the fiducial group luminosity $L_f(a)$, defined so
that at each expansion $90\%$ of the total mass is contained in groups with
luminosities $L \le L_f$.  The solid line shows the expected self-similar
evolution, normalized to the average of the two high-resolution experiments
for $\log a \in [-0.6, -0.2]$ as in Figure \ref{Mscale.fig}.}
\label{Lscale.fig}
\end{figure}
Figure \ref{Lscale.fig} shows the scaling of a fiducial group luminosity,
defined in much the same fashion as the fiducial temperature shown in
Figure \ref{Tscale.fig}.  For each group the total luminosity is defined as
$L = \int \rho^2 T^{1/2} ~dV$, which becomes a sum over the member SPH
particles $L = \sum_j m_j \rho_j T_j^{1/2}$.  As with the temperature, we
sort the groups in order of increasing luminosity and find the luminosity
$L_f$ such that $90\%$ of the total baryon mass is contained in groups with
$L \le L_f$.  The solid line shows the expected $L \propto a^{1/2}$ scaling
fitted to the high-resolution data for $\log a \in [-0.6,-0.2]$.  It is
evident that the total group luminosity scales quite poorly.  There is also
a rather large resolution effect: the low-resolution TreeSPH luminosities
are fainter than those of the the high-resolution case by an order of
magnitude.  The poor scaling of the luminosity is somewhat disheartening,
as the luminosity function is a basic, observationally testable prediction
that we would like to obtain from these sorts of simulations.  We must also
ask {\em why} the luminosity is scaling so poorly: we already know that the
mass and temperature scale reasonably, which suggests that the gas density
is the culprit.

\begin{figure}[htbp]
\plotone{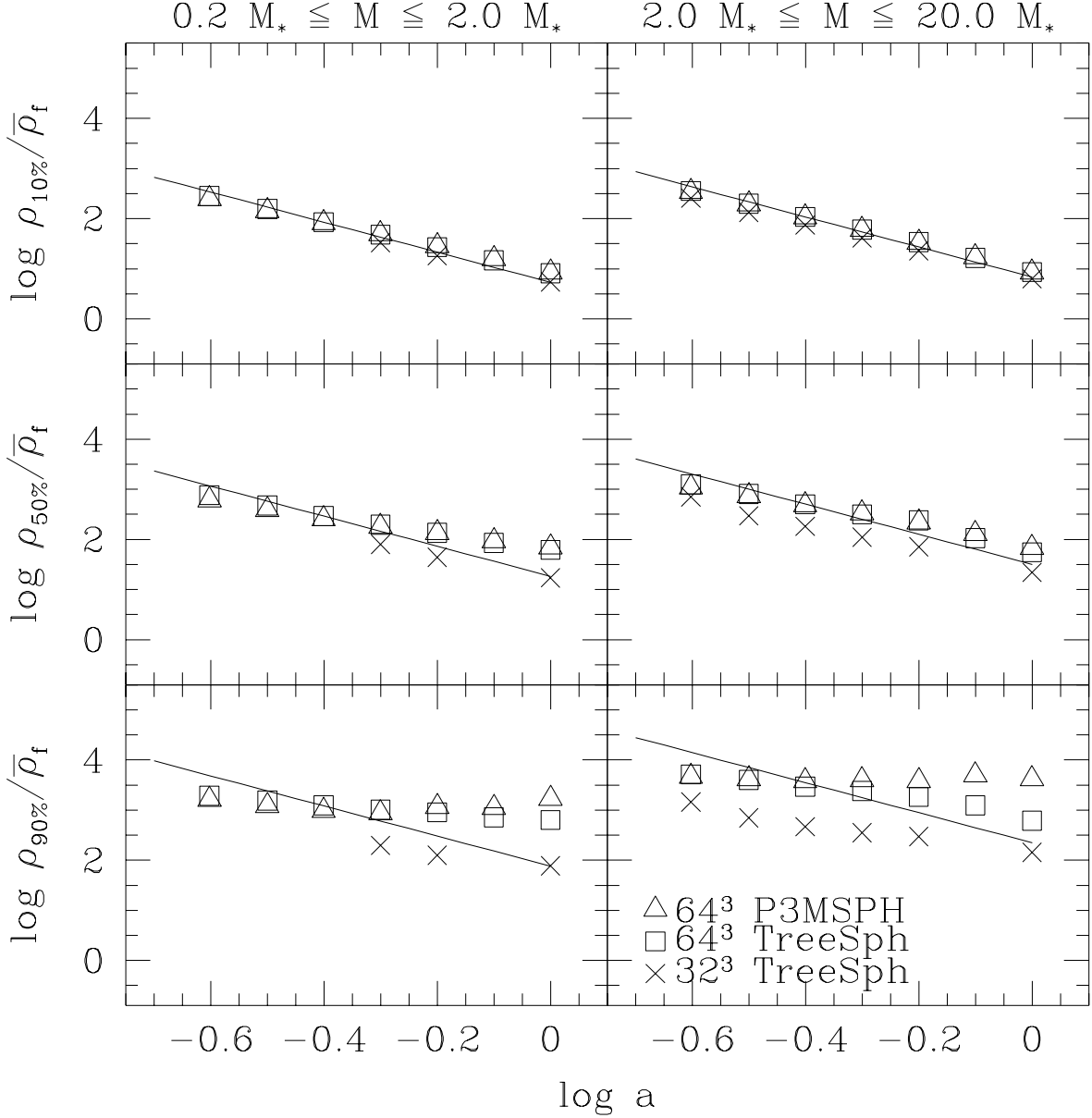}
\caption{Evolution of the group percentile gas density $\rho_{x\%}(a)$.
$\rho_{x\%}$ is defined as an average percentile gas density for groups in
a given mass range, where $\rho_{x\%}^i$ for each group $i$ is the density
such that $x\%$ of the baryon mass in the group is at densities $\rho \le
\rho_{x\%}^i$.  The left column shows the results for averaging over groups
in the mass range $0.2~M_* \le M \le 2.0~M_*$, while the right column
averages over groups in the range $2.0~M_* \le M \le 20.0~M_*$.  The top
row shows $\rho_{10\%}$, the middle $\rho_{50\%}$, and the bottom
$\rho_{90\%}$, progressively sampling from the outskirts of each group
inward as we go to higher $x$.  The solid lines show the expected
self-similar evolution, normalized to the average of the two
high-resolution experiments for $\log a \in [-0.6, -0.2]$.}
\label{rhoscale.fig}
\end{figure}
Figure \ref{rhoscale.fig} tests the scaling of the group gas density,
though in a slightly different fashion than was used in the previous tests
for mass, temperature, and luminosity.  The density needs to be treated
distinctly because the groups are effectively selected by their
overdensities, so in a sense the density should scale by construction.  In
order to generate this figure, we first calculate percentile densities
$\rho_{x\%}$ for each group, such that $x\%$ of the SPH particles in the
group are at densities $\rho_j \le \rho_{x\%}$.  We then take all of the
groups in a given mass range and find the average value of $\rho_{x\%}$,
where the average is weighted by the group mass.  We also impose the
additional constraint that a group must contain more than 32 particles to
be considered -- this is why the low-resolution experiment is missing from
the early expansions in the low mass range.  In Figure \ref{rhoscale.fig}
we plot $\rho_{10\%}$, $\rho_{50\%}$, and $\rho_{90\%}$, progressively
probing from the outskirts of each group inward.  We also consider two mass
ranges for selecting the groups: $M \in [0.2 ~M_*, 2.0 ~M_*]$ and $M \in
[2.0 ~M_*, 20.0 ~M_*]$, representing the low and high ends of mass range
probed by the simulations (see Figure \ref {MassRes.fig}).  As before, the
solid lines show the self-similar evolution law normalized to the
high-resolution experiments for $\log a \in [-0.6,-0.2]$.

Several trends are immediately obvious in the evolution of $\rho$.  The
density follows self-similar scaling best for low density cuts (such as
$\rho_{10\%}$), suggesting that the outer regions of the groups scale more
effectively than the inner, high-density cores.  The different simulations
also agree most closely among themselves for the lower density cuts.  For
instance, although $\rho_{10\%}$ agrees well for the two high-resolution
experiments, the core ($\rho_{90\%}$) densities differ by $a=a_f$.  This
difference, combined with the (physically related) temperature inversion in
the P3MSPH simulation, is the underlying cause for the divergence in the
emission weighted temperatures between the two high-resolution experiments,
noted in Figure \ref{Tscale.fig}.  There is a tendency for the high mass
range to scale better than the low mass range, though even for the higher
(and presumably better resolved) mass range, $\rho_{90\%}$ scales quite
poorly.  The scaling behavior seen in Figure \ref{rhoscale.fig} is not
surprising, given that collapsed objects have internal density profiles
extending to (potentially) arbitrarily high values.  Finite simulations
effectively smooth the density field of the exact solution on a fixed
Lagrangian scale ($\sim 32$ particles or so in this case), and the
existence of this fixed filter is responsible for the poor scaling at high
density contrast.  A simple way to think about this is that, in barely
resolved objects (\ie, at early times within a given mass bin),
$\rho_{90\%} \simeq \rho_{10\%}$ because both densities enclose an amount
of mass comparable to the filter scale.  Later, when the number of
particles in a group is $N \gg 32$, one expects $\rho_{90\%} \gg
\rho_{10\%}$ because much of the density profile is now ``exposed'' above
the Lagrangian filter.

As an aside, we note that the averages shown in Figure \ref{rhoscale.fig}
include only objects that contain more than 32 particles and are thus
nominally resolved.  For such nominally resolved structures, one could
argue that the regions probed by $\rho_{50\%}$ and $\rho_{90\%}$ contain
significantly fewer particles, and thus are not resolved.  However, the
results do not change substantially if we repeat the measurements of Figure
\ref{rhoscale.fig} while restricting ourselves to objects with more than 64
and more than 320 particles, so that $\rho_{50\%}$ and $\rho_{90\%}$,
respectively, are also resolved.

\begin{figure}[htbp]
\plotone{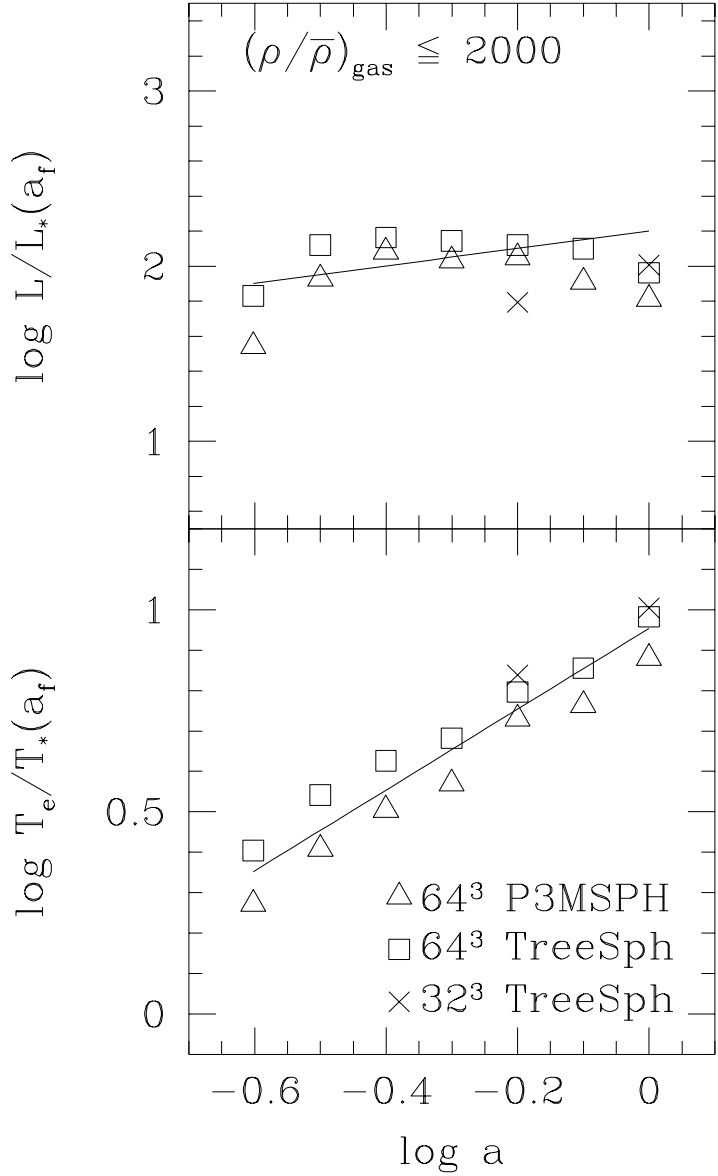}
\caption{Evolution of the group luminosity $L_f(a)$ (upper panel) and
emission weighted temperature (lower panel), as in Figure
\protect\ref{Lscale.fig} and the right panel of Figure
\protect\ref{Tscale.fig}, respectively.  In this figure, however,
only particles with densities $\rho/\bar{\rho} \le 2000$ are allowed to
contribute, so that we are effectively sampling in a ``shell'' surrounding
the core of each object.}
\label{LTscale2000.fig}
\end{figure}
Both the total luminosity and the emission weighted temperature of a group
are dominated by the high-density core, so the results of Figure
\ref{rhoscale.fig} suggest that the poor scaling of the luminosity noted
previously is likely due to the poor representation of the core density.
However, the gas density in moderate to low density regions in the groups
does scale well, so we might hope that the luminosity in these restricted
regions will more closely follow self-similar scaling.  Although the
emission weighted temperatures shown in Figure \ref{Tscale.fig} do scale
reasonably, we might expect that the scaling and agreement for these
temperatures will improve if we restrict ourselves to the regimes where the
density scales effectively.  In Figure \ref{LTscale2000.fig} we recalculate
the $90\%$ luminosity and emission weighted temperature evolution, allowing
only mass points that fall below an upper density cutoff of
$\rho/\bar{\rho} \le 2000$ to contribute for each group.  The luminosity
scaling (in the upper panel) is still not as expected, but there is
definite improvement compared with Figure \ref{Lscale.fig}, particularly at
late expansions.  The high-resolution experiments show far less evolution
of the luminosity with expansion, though the fit to the expected $L \propto
a^{1/2}$ scaling is still mediocre.  It is also notable that the scatter
between the different simulations is considerably reduced.  By the final
expansion even the low and high-resolution experiments agree, in stark
contrast with Figure \ref{Lscale.fig}.  The scaling of the emission
weighted temperatures (in the lower panel) also improves with the exclusion
of the cores, particularly in the P3MSPH experiment at late times.  This
has a lesser effect on the mass weighted temperatures (not shown here),
though, since the mass weighted temperature is not as strongly skewed
toward the highest density material as either the luminosity or the
emission weighted temperature.  However, the correspondence between the
mass weighted temperatures in the $64^3$ TreeSPH and P3MSPH simulations
does improve markedly with the exclusion of the cores, suggesting that most
of the difference in the temperature structures of these objects (such as
noted in Figure
\ref{Tscale.fig}) is restricted to their innermost, high-density regions.

\subsection{Group Temperatures and Luminosities}
\begin{figure*}[htbp]
\begin{minipage}[t]{0.45\hsize}
\plotone{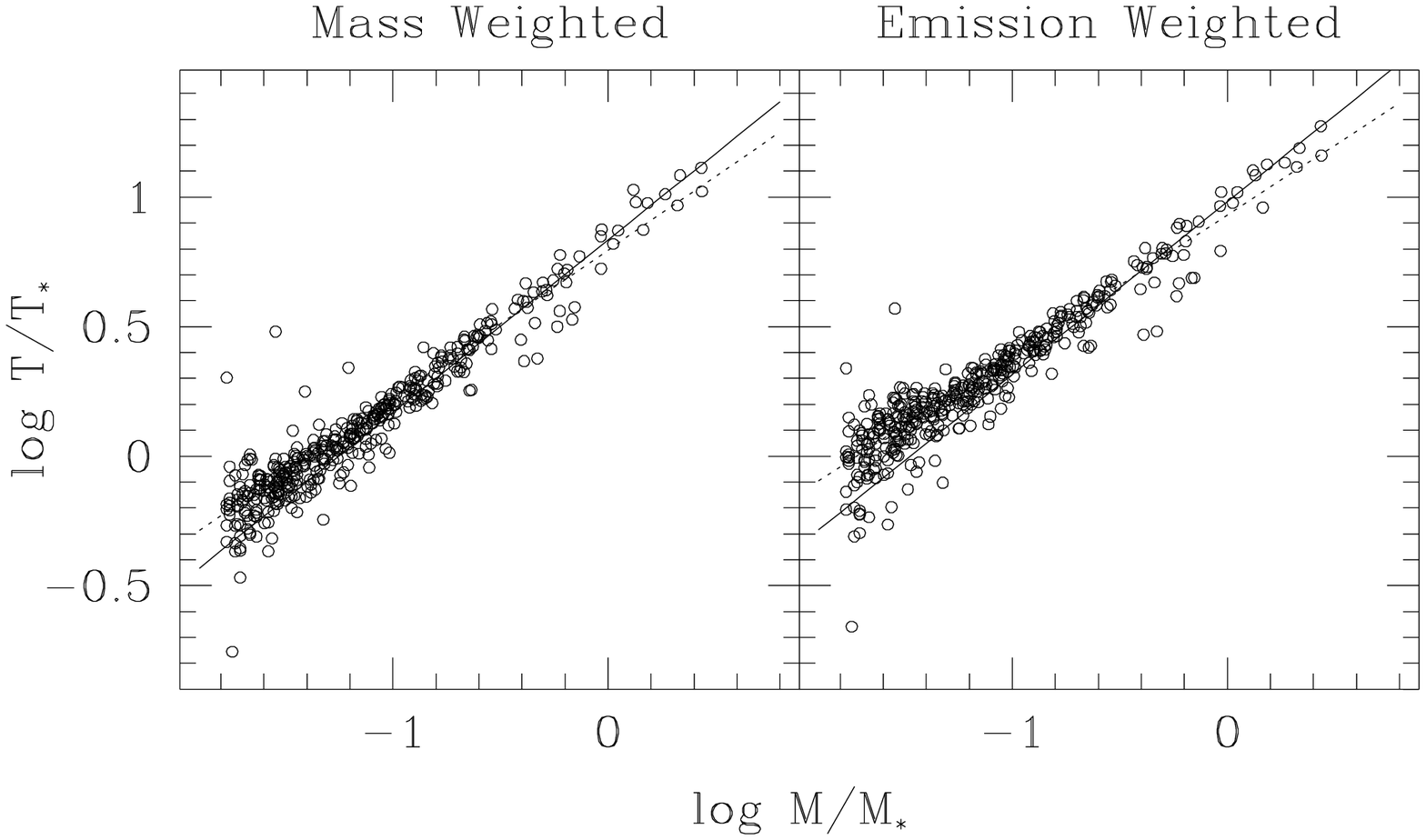}
\end{minipage}
\hfill
\begin{minipage}[t]{0.45\hsize}
\plotone{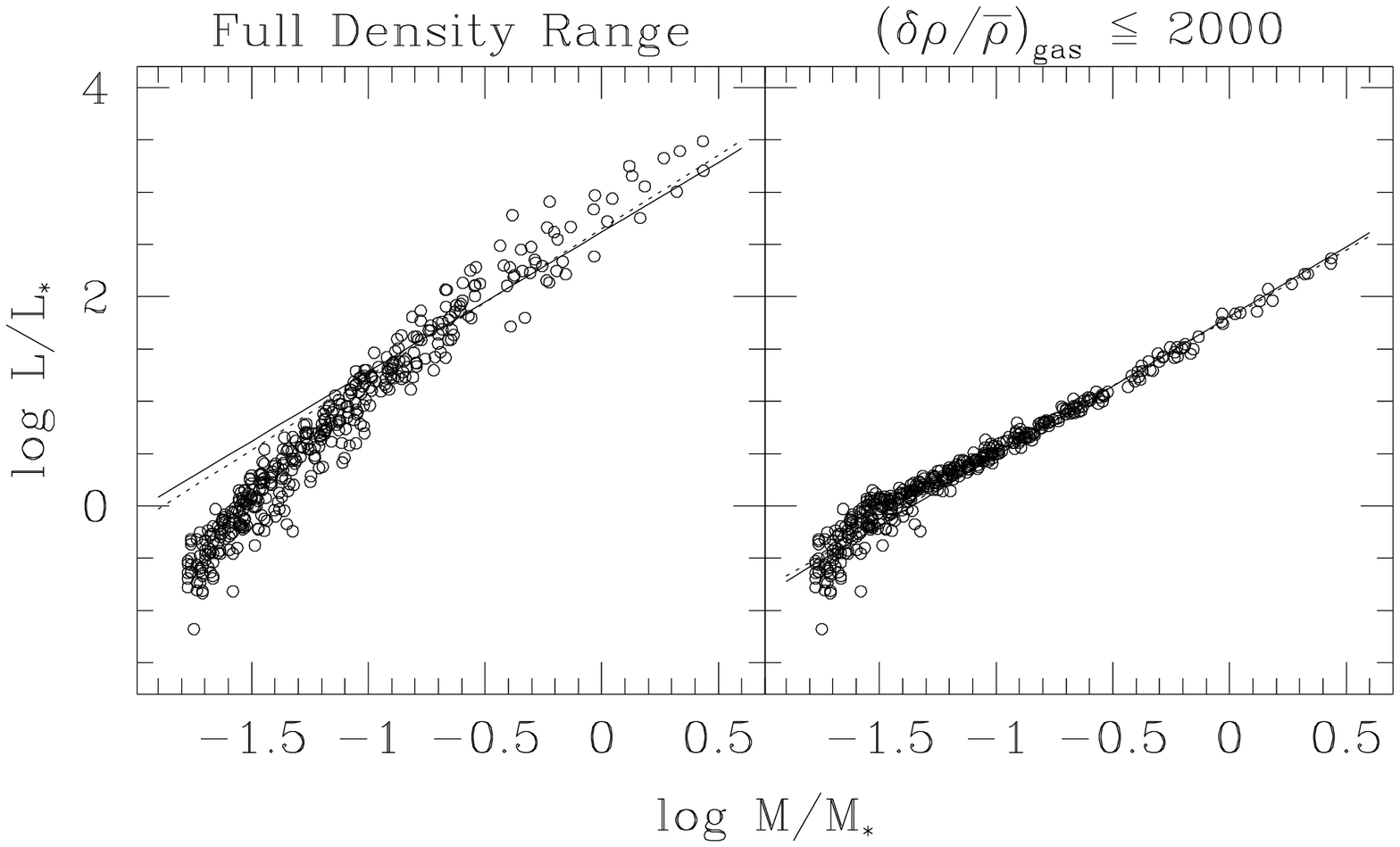}
\end{minipage}
\caption{As a function of individual group masses: (a) mass and emission
weighted temperatures; (b) luminosities for the $64^3$ TreeSPH simulation at
the final expansion, $a = a_f \equiv 1$.  The solid lines show the fit for
the scalings expected in the case of self-similar cluster structure, $T
\propto M^{2/3}$ and $L \propto M^{4/3}$.  The dotted lines in part (a) show
the best fit power-laws $T_m \propto M^{0.6 \pm 0.1}$ and $T_e \propto
M^{0.5 \pm 0.1}$, and in part (b) (left panel) total luminosities
$\Sub{L}{tot} \propto M^{1.4 \pm 0.1}$ and (right panel) ``shell''
luminosities $\Sub{L}{sh} \propto M^{1.3 \pm 0.1}$.  Only groups with more
than 32 particles are shown.}
\label{MTL64.fig}
\end{figure*}
X-ray observations of hot gas in galaxy clusters probe the temperature and
luminosity of these objects (see, for instance, Edge \etal\ 1990; Henry
\etal\ 1995; Burns \etal\ 1996).  In Figure \ref{MTL64.fig} we plot (a) the
temperature (both mass and emission weighted) and (b) the luminosity (total
and in shells with $\rho/\bar{\rho} \le 2000$) for each group identified in
the high-resolution TreeSPH experiment at $a=a_f$ against its baryon mass.
We include only resolved groups (\ie, those with more than 32 particles).
It is evident that the temperature and luminosity of a group are well
correlated with its mass, and in fact both follow relatively tight
power-law relations.  If we make the assumption that at any given time the
internal structure of groups with differing masses should be similar (\ie,
the density profile of a 0.5 $M_*$ group is a scaled version of that found
in an $M_*$ group), then simple scaling arguments predict $T \propto
M^{2/3}$ and $L \propto M^{4/3}$.  Kaiser (1990) uses this sort of
reasoning to make a detailed comparison between observed X-ray luminosity
functions and predictions based on self-similar models in combination with
the Press-Schechter mass function.  It is worth emphasizing, though, that
while self-similarity rigorously predicts the temporal evolution of
statistical distributions, $f(M,T,\ldots,t_1) \to f(M,T,\ldots,t_2)$, it
does not necessarily tell us about the detailed structure or arrangements
of objects at any given time.  While we know from our scale-free condition
that an $M_*$ group at time $t_1$ must be similar to an $M_*$ group at
$t_2$, we do not necessarily know that a 0.5 $M_*$ group is similar to an
$M_*$ group.  However, recent N-body experiments suggest that collapsed
objects built through hierarchical structure formation do tend toward a
universal density profile (Navarro, Frenk, \& White 1995, 1996, 1997; Cole
\& Lacey 1996) over roughly two orders of magnitude in mass, though these
studies also find some evidence that low mass halos tend to be denser.
Given these results, it seems reasonable to adopt the assumption that
objects will have similar density profiles over our limited mass range,
thus implying the relations $T \propto M^{2/3}$ and $L \propto M^{4/3}$.

\begin{figure}[htbp]
\plotone{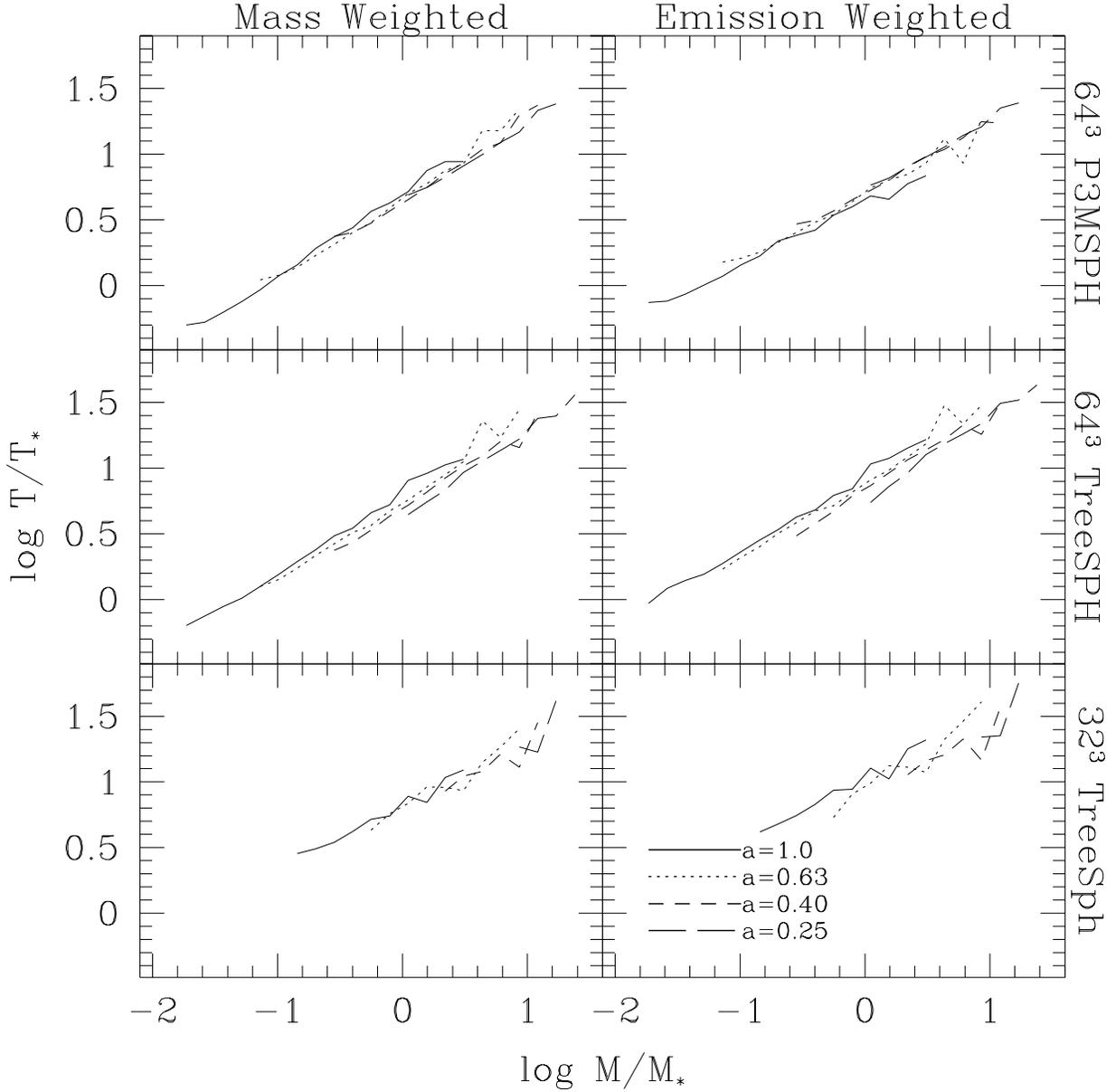}
\caption{Mass vs.\ temperature relations at expansions $a$ = 0.25,
0.4, 0.63, and 1.0.  At each expansion, the groups are binned by mass, the
average of the group temperatures (individual mass weighted group
temperatures in the left column, emission weighted in the right) is
calculated, and the resulting curves are scaled assuming self-similarity to
expansion $a = a_f \equiv 1$.  Only groups with more than 32 particles
contribute to the averages.}
\label{MT.fig}
\end{figure}

\begin{figure}[htbp]
\plotone{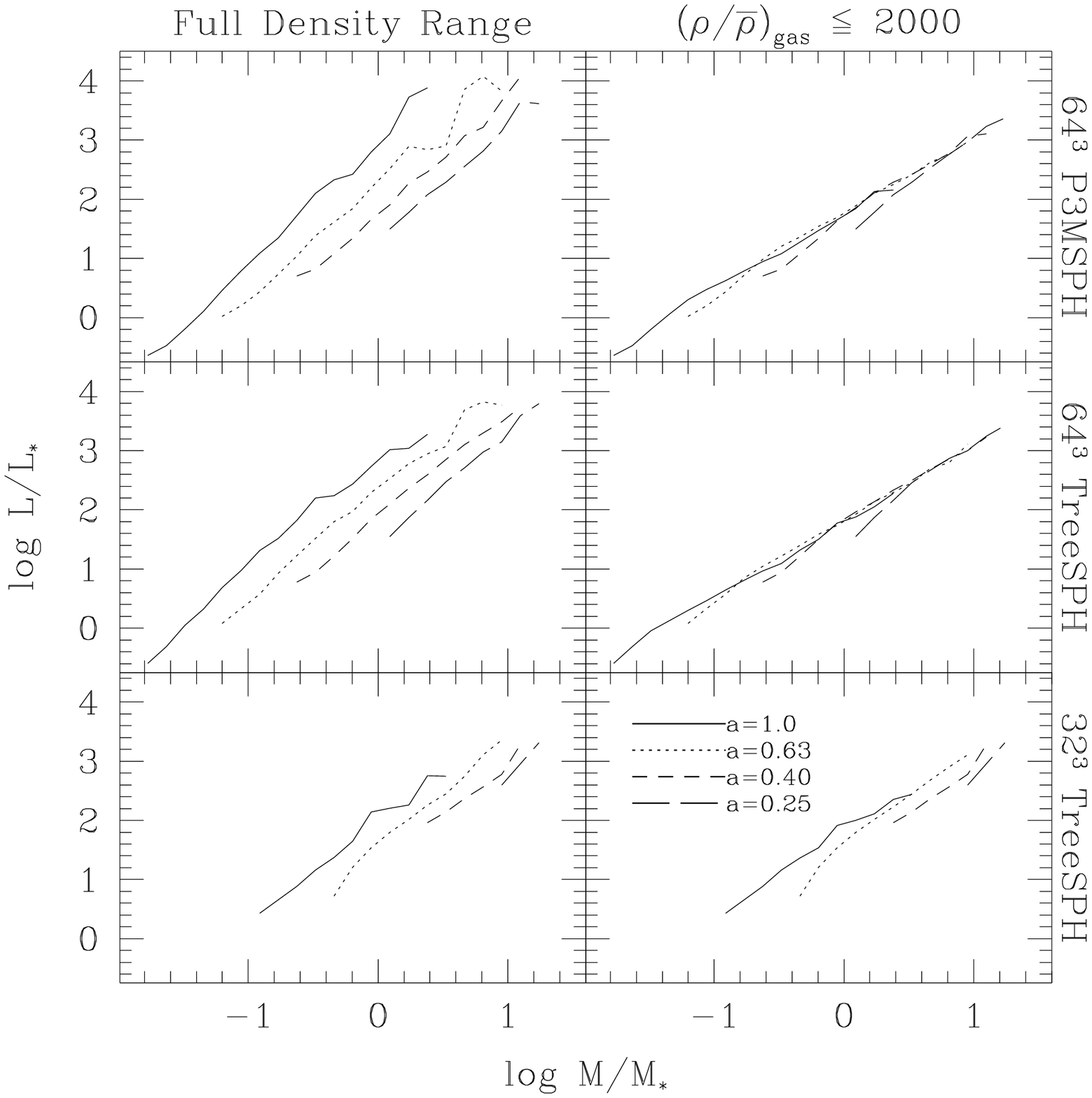}
\caption{Mass vs.\ luminosity relations at expansions $a$ = 0.25, 0.4,
0.63, and 1.0.  As with the temperatures in Figure \protect\ref{MT.fig}, at
each expansion, the groups are binned by mass, the average luminosity in
each bin is calculated, and the resulting curves are scaled assuming
self-similarity to expansion $a = a_f \equiv 1$.  The left column uses the
total luminosity for each object, while the curves in the right column only
use SPH particles in each group up to an upper density cutoff
$\rho/\bar{\rho} \le 2000$.  Only groups with more than 32 particles
contribute to the averages.}
\label{ML.fig}
\end{figure}

The solid lines in Figure \ref{MTL64.fig} show the expected scaling
relations normalized to the data, while the dotted lines show the
power-laws determined by linear least squares fitting to the average
relations in Figures \ref{MT.fig} and \ref{ML.fig}, described in more
detail below.  In Figure \ref{MTL64.fig}a there are two obvious, fairly low
mass groups lying well above the prevalent mass-temperature relation.
Close inspection of these outliers reveals them to be low mass objects in
the process of merging into larger structures.  We just happen to be
catching them in the process of being strongly shock heated and having
their gaseous halos stripped, though they still remain distinct objects as
defined by friends-of-friends.  The low-temperature outlier at the extreme
low-mass edge appears to represent the beginning of the spread for
unresolved groups, as the tight correlation between the masses and
temperatures of the objects becomes considerably looser as we go below the
nominal mass resolution of the experiment.  This point just happens to
barely make it above our low-mass cutoff.  More subtly, we can also see a
rough chain of groups just below and to the right of the general
mass-temperature trend.  Inspection of these objects reveals that they are
typically systems consisting of more than one (generally two to three)
visually distinct objects which are close enough to be linked together by
the friends-of-friends routine, but are not yet interacting significantly.
They therefore have more mass than their temperatures would ordinarily
imply.  Presumably, as these systems continue to coalesce they will shock
and thereby increase their temperatures so that they move up into the main
sequence of the mass-temperature measurements.  Turning to the luminosities
in Figure \ref{MTL64.fig}b, we can see that both the total and shell
luminosities are well correlated with the mass.  Given the results of
Figures \ref{rhoscale.fig} and \ref{LTscale2000.fig}, it is not surprising
that the shell luminosities form a better power-law of the mass once we
exclude the poorly resolved high-density cores.

Based on the tight correlation of the individual group masses with their
temperatures and luminosities, it seems reasonable to construct average
group mass-temperature and mass-luminosity relations.  Figure \ref{MT.fig}
displays the average mass-temperature relation at four different expansions
for each simulation.  Each of the curves is scaled assuming self-similarity
to $a = a_f$, so theoretically the curves should join up and form one
continuous sequence.  We can see that the mass-temperature relations do in
fact follow a reasonably continuous power-law.  A linear least-squares fit
to these relations finds for the mass weighted temperatures $T_m \propto
M^{0.6 \pm 0.1}$ ($64^3$ P3MSPH), $T_m \propto M^{0.6 \pm 0.1}$ ($64^3$
TreeSPH), and $T_m \propto M^{0.5 \pm 0.2}$ ($32^3$ TreeSPH).  For the
emission weighted temperatures we find $T_e \propto M^{0.5 \pm 0.2}$
($64^3$ P3MSPH), $T_e \propto M^{0.5 \pm 0.1}$ ($64^3$ TreeSPH), and $T_e
\propto M^{0.5 \pm 0.2}$ ($32^3$ TreeSPH).  Note that the measured best fit
mass weighted temperature relations are within one sigma of the relation,
$T \propto M^{2/3}$.  Due to resolution effects that will be elaborated in
\S \ref{DirectComp.sec}, we expect the temperature to be overestimated for
low-mass and presumably less resolved systems. This differential resolution
effect pushes the measured $T(M)$ relation to shallower slopes, so we
should take our measurements to represent a lower limit on the ``true''
(\ie, infinite resolution) value.  In other words, if we parameterize the
mass-temperature relation as $T \propto M^{\alpha_T}$ (where $T$ could be
either $T_m$ or $T_e$), then these experiments imply $\alpha_T \gtrsim
0.6$.

In Figure \ref{ML.fig} we consider the overall mass-luminosity relations.
In the left column we use all the particles in a group to define its
luminosity, while the right column plots the shell luminosities using only
particles with densities $\rho/\bar{\rho} \le 2000$.  As in Figure
\ref{MT.fig}, each expansion is scaled assuming self-similarity to
$a=a_f$, so we expect the curves to form a continuous relation.  Clearly
the total mass-luminosity relations in the left column fail this test,
while the shell luminosities in the right column do form a continuous
sequence.  Qualitatively, it appears that including the unresolved cores in
the total luminosity pushes the apparent power-law to a steeper slope,
presumably because the cores of higher-mass objects are better resolved.
We again use linear least-squares to fit a power-law to the mass-luminosity
relations: for the total luminosities in the left column we find $L \propto
M^{1.5 \pm 0.1}$ ($64^3$ P3MSPH), $L \propto M^{1.4 \pm 0.1}$ ($64^3$
TreeSPH), and $L \propto M^{1.2 \pm 0.2}$ for $32^3$ TreeSPH.  Measuring
the shell luminosities from the right hand column we find $L \propto M^{1.3
\pm 0.1}$ for $64^3$ TreeSPH and P3MSPH, and $L \propto M^{1.2 \pm 0.2}$ in
$32^3$ TreeSPH.  As we might expect from Figures \ref{Lscale.fig} and
\ref{LTscale2000.fig}, the scatter in the mass-luminosity relation tightens
substantially when we exclude the high-density cores.  However, considering
the mediocre scaling of the shell luminosities in Figure
\ref{LTscale2000.fig}, it is still somewhat surprising that the luminosity
relations from the different expansions link up as well as they do.  The
mass scaling test in Figure \ref{Mscale.fig} indicates that the group mass
distribution at the different expansions scales well, while the luminosity
scaling in Figure \ref{LTscale2000.fig} shows that we tend to overestimate
the true group luminosities at early expansions.  We would therefore
predict that the relation between $L$ and $M$ for an infinite resolution
realization might be shallower than what we measure, so if $L \propto
M^{\alpha_L}$ these results give us an upper limit $\alpha_L \lesssim 1.3$.

The mass-temperature and mass-luminosity relations have also been examined
by Navarro, Frenk, \& White (1995), who present a hydrodynamical study of
X-ray clusters selected from a large-scale CDM simulation.  Rather than
include hydrodynamics for every object in their volume, these authors
select a few interesting cluster scale structures from a large volume,
collisionless CDM simulation.  They then resimulate the evolution of these
selected objects at high resolution, including a baryonic component.  The
advantage of this approach is that they can simulate the development of
cluster-mass structures within a very large simulation volume (180 \himpc),
thereby correctly representing the long-wavelength gravitational power
while still achieving high resolution in these interesting regions.  The
major disadvantage is that they can only afford to study a few structures
within the volume at reasonable resolution, in this case eight.  Though
this low-number limit makes it difficult to pin down the power-law
relations as in Figures \ref{MTL64.fig}a and b, they find that their
measured mass, temperature, and luminosity relations are consistent with
the scalings expected for self-similar cluster structure, $T \propto
M^{2/3}$ and $L \propto M^{4/3}$.

\begin{figure}[htbp]
\plotone{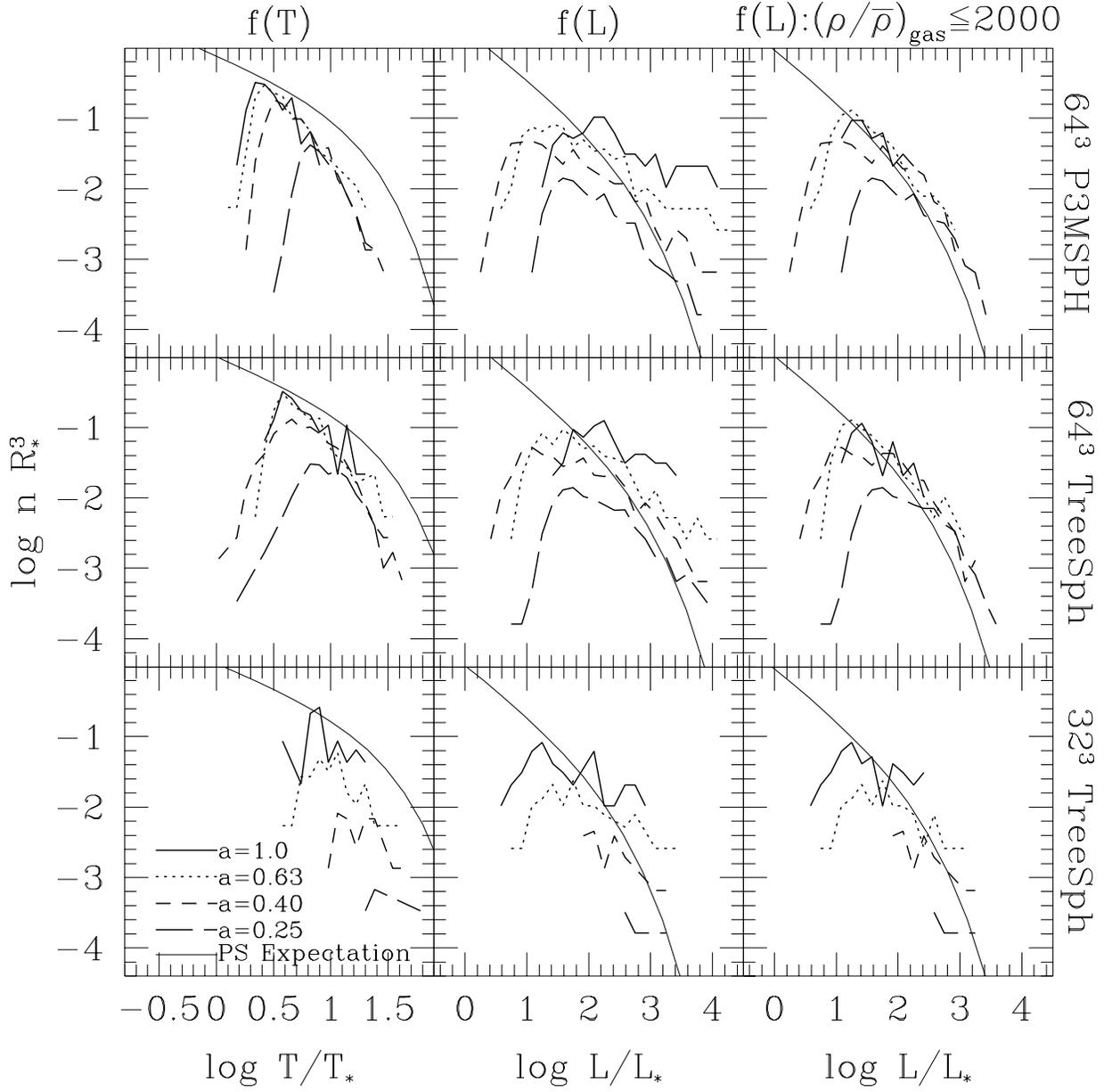}
\caption{The differential number distribution functions of emission
weighted group temperatures $f(T)$ and luminosities $f(L)$ for expansions
$a$ = 0.25, 0.4, 0.63, and 1.0.  The number densities $n$ are multiplied by
$R_*^3$ so that the scaled predictions are independent of epoch.  All
quantities are scaled self-similarly to $a = a_f \equiv 1$.  The left
column shows the distribution of emission weighted temperatures $f(T)$, the
middle $f(L)$ using all particles in each group, and the right $f(L)$ when
the luminosities are computed using only particles with densities
$\rho/\bar{\rho} \le 2000$.  The thin solid lines show the Press-Schechter
prediction, which results from combining the PS mass distribution with the
scalings $T \propto M^{2/3}$ and $L \propto M^{4/3}$.}
\label{fTL.fig}
\end{figure}
In addition to examining the scaling of fiducial or mean temperatures and
luminosities, we can study the evolution of their full distribution
functions, as we have already done for the mass in Figure \ref{fM.fig}.
Figure \ref{fTL.fig} computes the number distribution function of the
groups by emission weighted temperature and luminosity, $f(T)$ and $f(L)$.
We use the emission weighted temperature because it is the more
observationally relevant quantity, despite the fact that the mass weighted
temperature demonstrates better scaling.  We measure the distributions at
four different expansions shifted self-similarly to $a=a_f$, so the curves
should overlap.  Note that the total number of $M_*$ groups in the
simulation volume falls with expansion, so it is necessary to normalize by
the number of nonlinear volumes contained within our simulation volume.  In
Figure \ref{fTL.fig} we therefore plot the number density $n$ multiplied by
$R_*^3$.  The thin solid lines show the prediction if we assume the
Press-Schechter mass function and use the power-laws $T \propto M^{2/3}$
and $L \propto M^{4/3}$.  We empirically determine the normalization of
these power-laws by fitting them to the average $T(M)$ and $L(M)$ curves
depicted in Figures \ref{MT.fig} and \ref{ML.fig}.  In this sense the PS
curves in Figures \ref{MT.fig} and \ref{ML.fig} are not true predictions as
was the case in the mass distributions of Figure \ref{fM.fig}.  However,
changing this empirically determined normalization will only slide the
resulting PS relations horizontally in these figures, so the PS curves can
be treated as a prediction to some extent.  The distributions of
temperatures (left column) and shell luminosities (right column) connect
reasonably, indicating that these distributions follow the expected
self-similar scaling fairly well.  However, the distributions of total
luminosities (middle column) at different expansions do not connect up, a
manifestation of poor scaling of the total luminosities.  The PS prediction
for $f(M)$ agrees well with the numerical results in Figure \ref{fM.fig},
and the power-law relations connecting $M \to T$ and $M \to L$ are quite
tight, so not surprisingly we also find reasonable agreement between the PS
prediction and the measured results for $f(T)$ and $f(L)$.  However, the PS
prediction for $f(T)$ is somewhat shallower than the numerical results, so
the two disagree on the number of high-temperature objects.  Similarly, the
PS predicted $f(L)$ curve is steeper than the numerical measurements,
leading to a less significant but still noticeable discrepancy in the
number of highly luminous objects.  Due to the poor scaling of the
numerical distributions for the total luminosities, we cannot draw any firm
conclusions about that comparison.  Additionally, since we do not know what
the ``true'' distributions of $f(T)$ and $f(L)$ should be, we do not
necessarily know whether the PS or numerical results are closer to reality.
It is nevertheless interesting to see how these very different approaches
agree or differ from one another.

\section{Direct Comparison of the Simulations}
\label{DirectComp.sec}
In the previous sections, we found that the simulations demonstrate
reasonable self-similar scaling, so long as we are careful to take into
account the numerical limitations of each experiment.  In addition to the
scaling tests we have considered thus far, it is also interesting to ask
whether simulations using different resolutions and implementations of SPH
get the same answer, and if not, how and why they differ from one another.
In several of the previous figures we can see evidence of numerical
artifacts, even in objects that we would consider to be reasonably
well-resolved.  For instance, in the mass scaling test of Figure
\ref{Mscale.fig} there is a trend for the baryon mass in collapsed objects
to increase with improving resolution, even though the dark matter masses
appear to be stable.  Likewise, the total luminosities and central
densities of the collapsed structures are also resolution dependent.

Since each simulation is based on identical initial conditions, it is
possible to identify the same objects in each realization.  We use the
algorithm described in Weinberg \etal\ (1997) to match corresponding
objects in pairs of simulations.  Under this scheme, groups in each
simulation are first sorted in decreasing order of mass.  Then the first
(most massive) group from the first simulation is matched to the most
massive group within $\Sub{L}{box}/55.55$ that remains unmatched in the
second.  This process is repeated for progressively less massive objects
until the end of the list is reached.  Our end results are insensitive to
changes of the matching distance by factors of at least 4.  At this point
one might well ask why we need to use such a matching algorithm at all in
order to identify corresponding structures in supposedly identical
simulations.  Theoretically, of course, given perfect experiments we could
make unambiguous one-to-one matches of objects between the simulations.
However, when comparing the low- and high-resolution simulations, for
instance, it is possible that a single apparent structure in the
low-resolution simulation could be resolved into two or more individual
component structures in the high-resolution case.  Additionally, even when
comparing two different experiments at the same resolution, it is possible
that small numerical errors (the details of which presumably differ between
different codes) could put particles on different initial orbits during the
transition from linear to quasi-linear evolution.  These differences can
then be further amplified in the fully nonlinear regime, characteristic of
the structures we examine.  An edge-sensitive detection scheme such as
friends-of-friends does not select exactly the same population of objects
even though the particle distributions in the simulations are similar.  Our
matching procedure provides a simple and objective way to circumvent these
minor differences.

We use this procedure to match objects between $64^3$ P3MSPH and $64^3$
TreeSPH, and between $64^3$ TreeSPH and $32^3$ TreeSPH.  Comparing the
high- and low-resolution TreeSPH experiments allows us to directly
investigate the effects of resolution, while comparing the two
high-resolution simulations should tell us something about the variation
introduced by differences in the SPH implementations.  Examples of
implementation differences include the choice of interpolation kernel
(Bi-cubic Spline for TreeSPH, Gaussian for P3MSPH), the precise form of the
artificial viscosity used, the method of updating the local SPH
smoothing/resolution scale (TreeSPH maintains a fixed number of neighbors
per particle, while P3MSPH uses an integration method based on the
continuity equation), the tradeoff between timestepping and resolution
(TreeSPH uses individual timesteps per particle and allows each particle to
adapt its timestep to local physical criteria, while P3MSPH uses a fixed
timestep for all particles and instead applies the timestep stability
criteria to control the local resolution), and so forth.  We will not
attempt to go into the details of all of these implementation choices here:
the interested reader is referred to the descriptions of TreeSPH in HK89
and KWH96, and P3MSPH in Evrard (1988).  Ideally, of course, these sorts of
numerical implementation issues should not affect the physical results at
all, and therefore it is interesting to investigate to what extent this is
true.

In this Section we consider only the results of the simulations at the
final expansion, $a=a_f$.  At this output time we have 1852, 1884, and 471
groups in the dark matter and 897, 1227, and 257 groups in the baryons for
the $64^3$ P3MSPH, $64^3$ TreeSPH, and $32^3$ TreeSPH simulations,
respectively.  Of these, we find matches for 1488 in the dark matter and
780 in the baryons between $64^3$ P3MSPH and $64^3$ TreeSPH.  We find
matches for 303 groups in the dark matter and 165 in the baryons for $32^3$
TreeSPH and $64^3$ TreeSPH.  The difference in the number of identified
groups (particularly the number of baryon groups in $64^3$ P3MSPH and
$64^3$ TreeSPH) is almost entirely confined to the low to unresolved mass
range of objects, and therefore should not be taken as highly significant.
For the sake of brevity, for the remainder of this section we will use 64P
to denote the $64^3$ P3MSPH simulation, 64T to denote $64^3$ TreeSPH, and
32T for the $32^3$ TreeSPH case.

\begin{figure}[htbp]
\epsscale{0.8}
\plotone{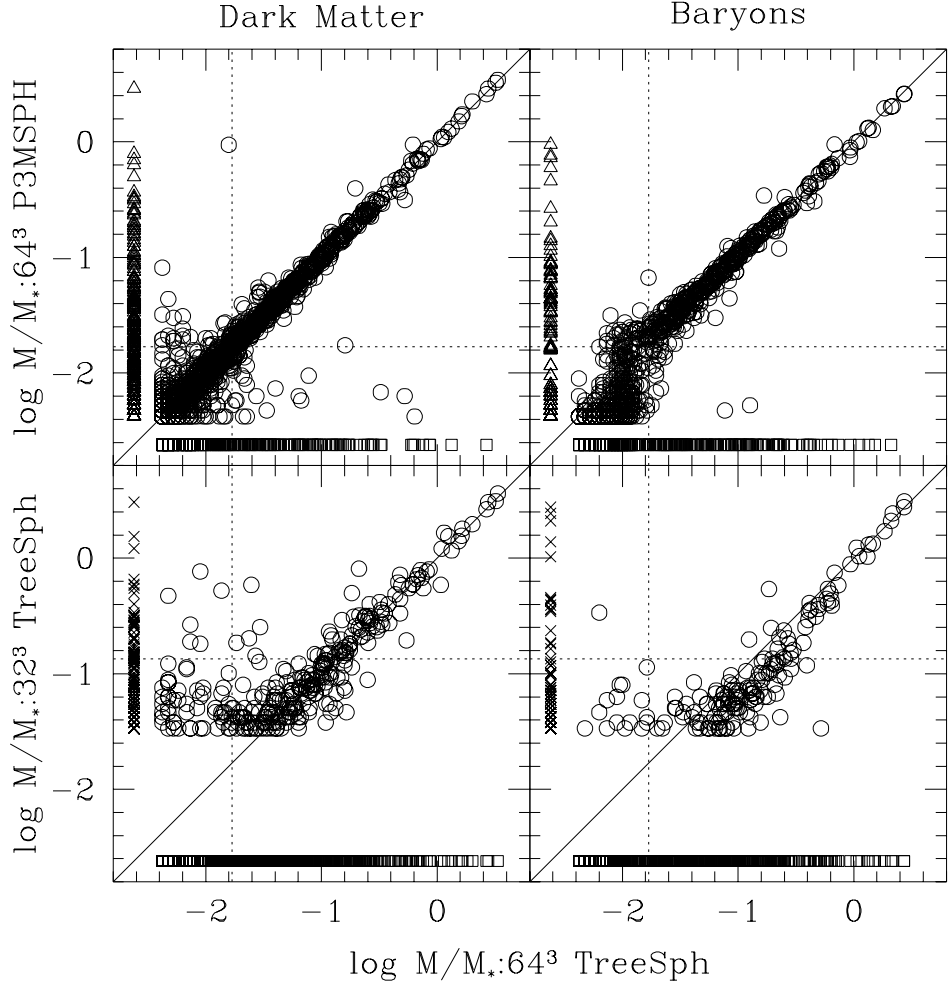}
\epsscale{1}
\caption{Group by group comparison of the group masses in dark matter (left
column) and baryons (right column), in different simulations at $a = a_f
\equiv 1$.  The top row compares $64^3$ P3MSPH to $64^3$ TreeSph; the
bottom row compares $32^3$ TreeSph to $64^3$ TreeSph.  In each panel the
circles represent groups which are matched between the simulations (see
text), while the points lying along the axes show groups which remain
unmatched.  The dotted lines show the 32 particle mass threshold.}
\label{MtoM.fig}
\end{figure}
Figure \ref{MtoM.fig} compares the group masses at $a=a_f$, where we plot
the 64P--64T comparison in the upper panels and the 32T--64T comparison in
the lower ones.  The open circles represent groups which have been matched
between the two simulations, while the differing point types along the axes
are groups which remain unmatched.  The dotted lines show the 32 particle
mass limit, representing our estimate of the mass resolution.  The dark
matter masses clearly match quite well for all of the simulations, with
almost all of the resolved groups falling nicely along the diagonal line.
This is the expected result, as it is well established that purely
collisionless N-body simulations converge once the nonlinear mass scale is
resolved.  Examining the baryon masses in the 32T--64T comparison, the
largest mass groups appear to match well.  However, we can see some slight
evidence for a resolution effect as we go to progressively smaller (and
therefore more poorly resolved) objects, in the sense that the baryon mass
is suppressed with poorer resolution.  The effect is quite small in this
Figure, and it is really only noticeable for objects up to 0.5 dex above
the 32 particle cutoff in the $32^3$ simulation.  Comparing the 64P and 64T
runs, the baryon masses for resolved objects concur very nicely.  Only
below the resolution limit do we see any discrepancy between these
experiments, at which point we know numerics dominate the results anyway.

\begin{figure}[htbp]
\epsscale{0.8}
\plotone{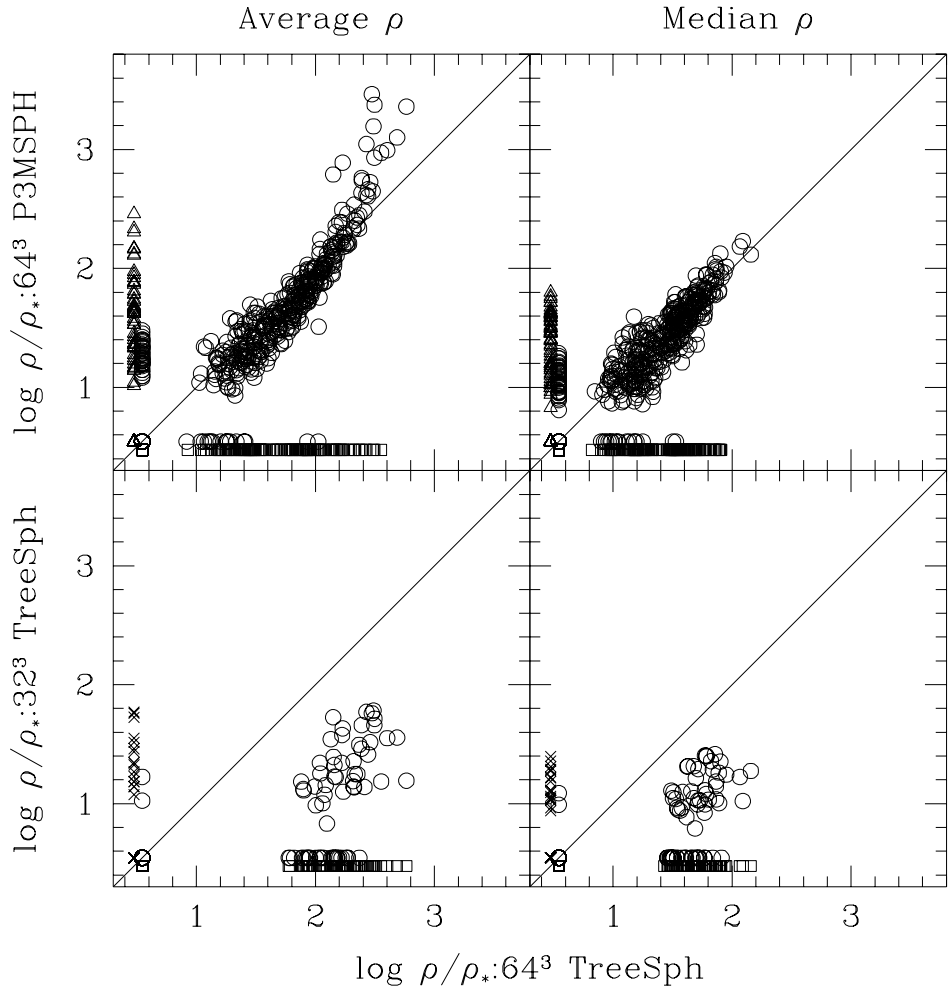}
\epsscale{1}
\caption{Group by group comparison of the gas density at $a = a_f \equiv
1$.  The left column uses the mass weighted average of the density, while
the right shows the median density.  The median density for each group is
defined to be the gas density such that $50\%$ of the SPH particles in the
group are at that density or lower, corresponding to the middle row of
Figure \protect\ref{rhoscale.fig}.  Only groups containing more than 32
particles are shown.}
\label{rhotorho.fig}
\end{figure}
In Figure \ref{rhotorho.fig} we directly compare the gas densities in the
groups, with the mass weighted average densities in the left column and
median densities in the right.  The median plotted here is the $50\%$ gas
density, corresponding to the middle row of Figure \ref{rhoscale.fig}.  As
anticipated, the gas density seems to be quite sensitive to the numerics.
The densities found in the 32T run always underestimate those found in the
higher resolution 64T experiment, with discrepancies ranging to nearly an
order of magnitude in the average density, and smaller discrepancies of
order 0.5 dex in the median.  The different behavior of average and median
densities is due to the fact that the mass averaged density of a group is
dominated by the high density gas in the unresolved core.  Comparing the
average density in the two high-resolution experiments we find that for
low to moderate mass objects the average density matches reasonably well.
However, the highest mass objects in P3MSPH have systematically higher
average densities than in TreeSPH.  These differences vanish in the median
density measurement, indicating that the variations in the average are
likely restricted to the high-density, unresolved cores.

The maximum density that an SPH simulation produces in an $M_*$ object at a
given epoch depends strongly on the number of particles and on the
numerical parameters that control the normalization of the smoothing
kernel.  This parameter can be specified in terms of the number of
particles within the finite domain of the kernel: $2 h_s$ for TREESPH, or
the $99\%$ power radius $2.38 h_s$ for the Gaussian used in P3MSPH, where
$h_s$ is the SPH smoothing length.  In the simulations reported here, the
former was set to 32 for TreeSPH, while the latter was 85 for P3MSPH.  It
is somewhat surprising, therefore, that the core densities of massive
objects are actually higher in the P3MSPH simulation.  The difference in
kernel normalization should act in the opposite direction, and in separate
checks we have found that within either code the central densities vary in
a systematic manner with kernel normalization in the expected direction
(lower densities with larger kernels).  We are unsure of the reason for the
observed difference in core densities between the two codes.  It may have a
dynamical origin related to the different ways the codes handle shocks, for
example, or it may be associated with more strictly numerical issues, such
as the way the codes enforce symmetry in the difference equations for
pairwise interactions.

\begin{figure}[htbp]
\plotone{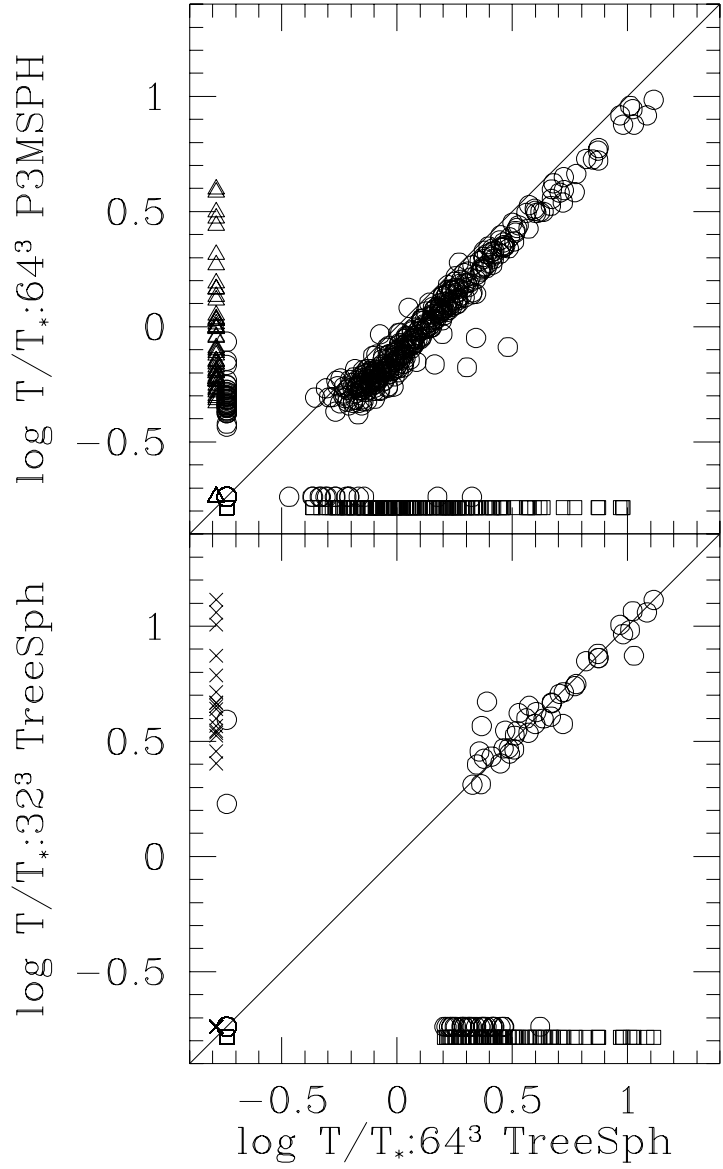}
\caption{Group by group comparison of the mass weighted group temperatures
at $a = a_f \equiv 1$.  Only groups containing more than 32 particles are
shown.}
\label{TtoT.fig}
\end{figure}
Figure \ref{TtoT.fig} plots the mass weighted temperature comparison.  The
fact that the group temperature scales well in Figure \ref{Tscale.fig}
suggests that the temperature is one of the more reliably measured
quantities, and this comparison bears that impression out.  Temperatures in
the 32T simulation very nearly match those of the high-resolution
experiment.  Similarly, the 64P and 64T temperatures correlate very
tightly, though the P3MSPH objects are systematically $25\%$ cooler than
their counterparts in TreeSPH.  Though not shown here, this distinction
between the 64P and 64T simulations almost entirely vanishes when we
exclude the high density core gas (defined as $\rho/\bar{\rho} \ge 2000)$,
implying that this discrepancy is again due to differences in the treatment
of the innermost, highly collapsed gas.  This difference is probably
related to the previously noted difference in the core densities: as the
core density increases, the temperature must drop if the core is to
maintain the same degree of pressure support (see, \eg, Shapiro \&
Struck-Marcel 1985 or Owen \& Villumsen 1997).  Alternatively, it is
possible that this is the result of the two codes converting kinetic energy
to thermal energy through the artificial viscosity in a slightly different
manner.  Though the total amount of energy being converted is roughly
equivalent, it could be that the details of how these two codes implement
the artificial viscosity result in differences in precisely where the
viscous interactions occur.

The slight discrepancy in temperatures between the P3MSPH and TreeSPH
experiments is not due to differences in the halo potentials of these
objects.  When we compare the velocity dispersions of individual objects
(in either the dark matter or the gas), the two TreeSPH experiments and the
P3MSPH experiment all match closely.  There is some indication that the
baryon velocity dispersion for high mass objects is higher in P3MSPH than
in TreeSPH at the few percent level, but this difference is insufficient to
explain the discrepancy in the temperatures.  The small temperature
difference we note in Figure \ref{TtoT.fig} does not reflect a significant
difference in the energy budgets of these objects: the potentials, velocity
dispersions, and gross energetics of the identifiable structures agree very
well between the different experiments, leaving the resolution or
implementation differences discussed above as the likely cause of the
slight temperature differences.  Though not shown here, the emission
weighted temperatures do not agree quite as well as the mass weighted: the
low-resolution 32T experiment tends to find emission weighted temperatures
hotter than 64T by roughly $25\%$, while the emission weighted temperatures
in 64P are cooler by up to a factor of 2.5 at the high mass end.  As we
would expect based on the evidence of Figures \ref{Tscale.fig} and
\ref{LTscale2000.fig}, these differences are reduced if we exclude the high
density core gas, but they are not completely eliminated.  Over most of the
mass range, however, the differences in both mass weighted and emission
weighted temperatures are small, and the agreement between simulations and
the success of the scaling tests in \S \ref{SelfSim.sec} imply that these
numerical methods provide robust temperature predictions.

\begin{figure}[htbp]
\plotone{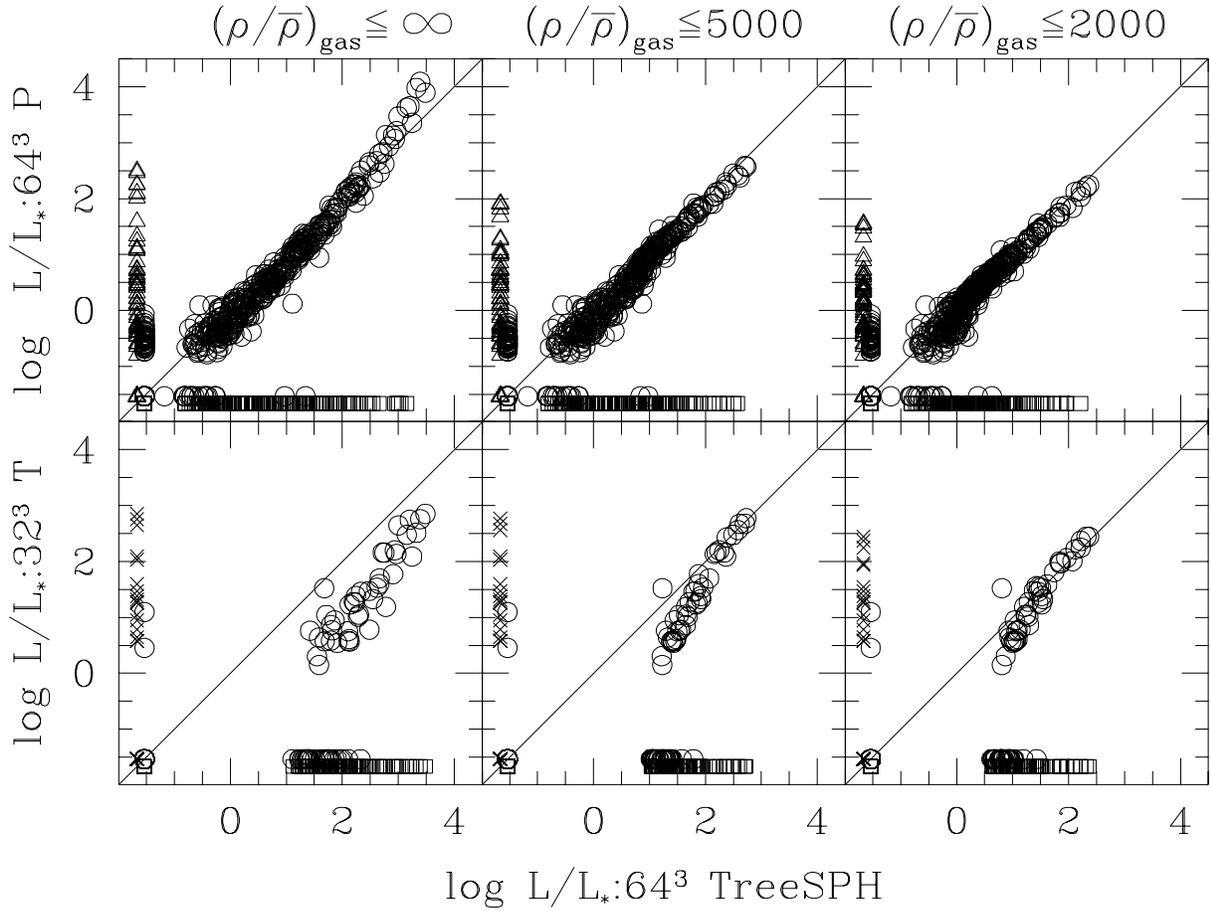}
\caption{Group by group comparison of the group luminosities at $a = a_f
\equiv 1$.  In the left column all particles in each group are used, in the
middle only particles with $\rho/\bar{\rho} \le 5000$, and in the right
$\rho/\bar{\rho} \le 2000$.  The top row shows the $64^3$ P3MSPH--$64^3$
TreeSPH comparison, and the bottom $32^3$ TreeSPH--$64^3$ TreeSPH.  Only
groups containing more than 32 particles are shown.}
\label{LtoL.fig}
\end{figure}
The scaling results in Figures \ref{Lscale.fig} and \ref{LTscale2000.fig}
suggest that the luminosity is very sensitive to numerical artifacts, as is
also evident in the direct comparison of Figure \ref{LtoL.fig}.  In this
Figure we impose three different upper thresholds on the gas density of
particles allowed to contribute to the luminosity: the left column shows
the total group luminosities using all particles from each group, the
middle column enforces a relatively high density cutoff of $\rho/\bar{\rho}
\le 5000$, and the right column uses a more restrictive upper density
threshold of $\rho/\bar{\rho} \le 2000$.  Clearly the groups in 32T tend to
underestimate the total luminosities compared with the 64T.  However, as we
impose upper limits on the density, the high mass groups begin to agree.
Furthermore, the more restrictive the upper limit we enforce, the greater
the mass range of groups which concur.  We see a similar trend in the
64P--64T comparison.  Based on the previous figures, it is not surprising
that the upper-end of the luminosity distribution differs between the two
high-resolution experiments.  However, once we impose upper density
cutoffs, the shell luminosities agree well, indicating that the luminosity
differences between the two experiments are restricted to the collapsed
cores of the groups.

Taken together, the trends noted in Figures \ref{MtoM.fig}--\ref{LtoL.fig}
reinforce the impression from the scaling tests: that numerical artifacts
in otherwise resolved objects are most evident in the central, highest
density regions.  The most direct effect of having limited resolution is
that we underestimate the central gas density, and to a lesser extent the
total baryon mass.  In agreement with previous investigations, we find that
the temperature is the most robust measurement, though there are small
numerical effects noted here as well.  Since the total luminosity is
dominated by the central, high-density gas, the density effect can, in some
regimes, cause drastic underestimates of the total luminosity.  The evident
stability of the mass weighted temperature is due to its fundamental
connection to a conserved quantity, energy.  Codes that conserve energy
locally and thermalize infall kinetic energy to the same degree must wind
up with similar mass weighted temperatures.  The distribution of that
thermal energy within collapsed objects is sensitive to a number of
details, such as the implementation of shock dissipation.  Differences in
these details apparently lead to the discrepancies found in the solutions
within the core regions.

Evidence for these kinds of resolution effects, particularly for the gas
density, is neither a novel result nor particular to SPH.  Kang \etal\
(1994b) perform a comparison of several cosmological hydro-codes (both SPH
and Eulerian-grid), and they also find that in collapsed objects the
central density can be quite resolution dependent, while the temperature
profile (being more flat-topped) is more reliably represented.  More
recently, Anninos \& Norman (1996) have studied the collapse and formation
of a typical galaxy-cluster sized object in an Einstein-de Sitter cosmology
($\Sub{\Omega}{bary} = 0.06$, $\Sub{\Omega}{dm} = 0.94$), using a nested
Eulerian grid code to simulate the hydrodynamics.  Even at the most refined
subgrid (effectively a $512^3$ grid covering the cluster), they do not find
convergence of the integrated luminosity or central density profile, while
the emission weighted temperature is more reliably represented.  In a study
of the effects of photoionization on structure formation, Weinberg \etal\
(1997) find that the cooling balance (and therefore the eventual collapsed
fraction) in marginally resolved structures can be strongly affected by the
numerical tendency to underestimate the gas density.  Owen \& Villumsen
(1997), in a survey of hierarchical structure formation scenarios using 2-D
SPH simulations, also find that in general the gas density distribution
does not converge with increasing numerical resolution.  However, they do
find that once a minimum temperature is imposed in the gas, the gas
distribution does converge if the corresponding Jeans mass is resolved.
The one study that runs counter to these trends is that of Navarro \etal\
(1995), mentioned previously.  In this case, the investigators find that
the properties of their X-ray clusters show evidence of convergence at
their highest resolutions.  Since they only simulate a small number of
objects, they are able to use much higher resolutions for these objects
than achieved here ($\approx 10^4$ SPH particles within $r_{200}$ for a
single cluster, where $r_{200}$ is the radius within which the mean
overdensity is $\delta \rho/\bar{\rho} = 200$, as opposed to $\approx
10^2-10^3$ particles within $r_{200}$ for typical objects in our
experiments).  It is possible that at these resolutions the cluster
properties really do converge, though this seems at odds with the results
found by Anninos \& Norman (1996) (whose highest resolution experiment
effectively has $\approx 6 \times 10^3$ Eulerian cells within $r_{200}$).

Comparing the two high-resolution simulations considered here, it appears
that the results of our experiments are relatively insensitive to the
minor implementation differences between TreeSPH and P3MSPH.  The only
significant distinction we note is that the high-mass objects in P3MSPH
tend to have higher core densities than their correspondents in TreeSPH.
However, we also find that these high-density cores are in fact unresolved,
so the results in these regions should not be taken very seriously anyway.
Once we restrict ourselves to regions that do seem to be effectively
resolved, the P3MSPH and TreeSPH experiments agree very closely.

\section{Summary}
\label{Summary.sec}
We analyze three 3-D hydrodynamical simulations evolved from scale-free
initial conditions in order to study self-similar evolution of hierarchical
structure formation including a gaseous component.  We consider only a
single model using an $n=-1$ power spectrum of density fluctuations.  We
evolve three independent simulations based on these identical initial
conditions: two high-resolution experiments performed with $64^3$ dark
matter and SPH particles -- one performed with TreeSPH and one with P3MSPH
-- and one low-resolution run performed with $32^3$ dark matter and SPH
particles under TreeSPH.  Because both the physics and the initial
conditions of these experiments are scale-free, we can make use of the
powerful prediction that these systems should evolve self-similarly over
time.  Using temporal self-similarity in combination with the known initial
power-spectrum index $n$, we know how characteristic quantities such as the
mass, temperature, and luminosity should evolve.  We test for self-similar
evolution in the simulations by identifying objects consisting of groups of
particles of a given overdensity at various times and examining how the
properties of these distributions of groups evolve.  Due to the mass
resolution limits of our experiments, a reasonable fraction of the total
expected mass distribution of structures is only accessible for roughly
$\Delta \log a \approx 0.4$, or about a factor of 2.5 in expansion.  During
this rather restricted interval we recover the expected scalings for the
mass and temperatures of the groups, while we find that the densities and
total luminosities scale much more poorly.  Upon further investigation, we
find that it is the central, high-density regions of the groups that are
causing the apparent poor scalings, and if we impose upper overdensity
cutoffs for the particles considered in each group (thereby restricting
ourselves to density ``shells'' around the core of each group), the
resulting densities and luminosities do scale reasonably.

The temperatures and luminosities of the groups show tight correlations
with their baryon masses, and we find that these relations are well
quantified by power-laws.  Restricting the luminosity contributions to
shells outside the core of each group greatly reduces the scatter in the
power-law relation between luminosity and mass, and it also tends to force
$L(M)$ toward a shallower slope.  Taking resolution effects into account,
we estimate that the measured power-law index for $T(M)$ is best viewed as
a lower limit, while that for $L(M)$ is an upper limit.  Representing these
power-laws as $T \propto M^{\alpha_T}$ and $L \propto M^{\alpha_L}$, we
empirically determine these exponents to be $\alpha_T \gtrsim 0.6 \pm 0.1$
and $\alpha_L \lesssim 1.3 \pm 0.1$.  These estimates are consistent with
the prediction $\alpha_T = 2/3$ and $\alpha_L = 4/3$, which results from
assuming that objects of differing masses have similar structures.

We find that the Press-Schechter prediction for the mass distribution
function matches the numerical results well for both the dark matter and
baryon masses, though PS tends to overpredict the amount of baryon mass
contained in objects at all scales.  This agrees with the result found in
purely collisionless studies such as Efstathiou \etal\ (1988), though we
confirm the agreement between the PS prediction and the numerical results
here for the baryonic component as well.  The agreement between the PS
prediction and the numerical results is all the more remarkable for the
fact that this is a prediction, not a fit, as there are no free parameters.
Since the power-law relations between the baryon mass and the group
temperature and luminosity are quite tight, we find that when we use these
relations to map the PS mass prediction to the temperature and luminosity
distributions it also matches the numerical results reasonably, though
there are some discrepancies at the high temperature/luminosity end.  The
procedure works best for the temperatures and for the ``shell''
luminosities, where we exclude the contributions to the luminosity from the
highest density regions in each object.  By contrast, the distribution of
total luminosities does not scale well to different times, and it does not
match the PS prediction.

The examples of temperature and luminosity distributions illustrate the
power of combining numerical simulations with self-similar arguments.
Self-similarity can tell us how to map the distribution of a quantity over
time, but it cannot predict the precise form or amplitude of that
distribution at any given time.  However, numerical simulations can compute
the distribution over a restricted range of expansions set by the
resolution limits.  Combining these two approaches, one can use numerical
simulations to calculate the detailed form of the distribution for
interesting quantities such as the temperature or luminosity function and
use self-similarity to scale this distribution to any point in time.

By direct comparison of the properties of groups in our different
experiments, we find that in general the obvious mass limitations outlined
in Figure \ref{MassRes.fig} are reasonable estimates of the mass and
expansion ranges we can probe.  However, we also find that the central
high-density cores of the collapsed objects are unresolved and therefore
dominated by numerical effects, even for objects which we would otherwise
predict to be adequately resolved.  The most strongly affected quantities
are the central gas density and luminosity, both of which tend to increase
with improving resolution.  The gas density can increase by up to an order
of magnitude when the SPH particle number is increased by a factor of
eight.  The most robust quantities are the group mass and temperature,
which we find to be only very weakly dependent on resolution.  Similar
numerical effects have been noted in previous studies for a variety of
hydrodynamical techniques (Kang \etal\ 1994b; Anninos \& Norman 1996;
Weinberg \etal\ 1997; Owen \& Villumsen 1997).

The fact that Efstathiou \etal\ (1988) identify the correct self-similar
scaling for collisionless N-body codes represents one of the great
successes for such techniques.  In this study we find that, under certain
restrictions, this self-similar behavior is still maintained when
hydrodynamical processes are added to such experiments, lending credence to
results based on such simulations.  However, there are several complex
numerical effects that come into play for these sorts of simulations
beyond those found in purely collisionless investigations, which must be
accounted for if the final results are to be believed.  Additionally, we
have omitted one important physical process in these experiments: radiative
cooling.  While radiative cooling is probably not important for the hot
intergalactic gas found in clusters and groups of galaxies (outside of
possible cooling flows in the cores of such objects), it is crucial to the
process of galaxy formation.  Understanding the process of galaxy formation
is a major goal in studies of cosmological structure formation today, and
many investigators are already using hydrodynamical simulations in order to
approach this problem.  Unfortunately, introducing a physically motivated
cooling law (such as that for a primordial H/He gas) violates the
scale-free requirement for self-similar evolution.  It is possible,
however, to construct artificial cooling laws that do maintain the
scale-free requirement for a given power-spectrum of initial density
fluctuations.  In future work we will use this approach to examine
self-similar behavior in simulations that incorporate radiative cooling,
which should have direct implications for numerical studies of galaxy
formation.

\acknowledgements
This work was supported by NASA grants NAG5-2882, NAG5-3111, NAG5-3820,
NAG5-2790, NAG5-4064, NAG5-3922, and NAG5-3525, and also by the NSF through
grant ASC 93-18185 and the Presidential Faculty Fellows Program.  A.E.E.
acknowledges the hospitality of the IAP, and support from the CNRS and CIES
of France, during a sabbatical stay.  We acknowledge computing support from
the San Diego Supercomputer Center and the Pittsburgh Supercomputing
Center.  This work was partially supported under the auspices of U.S. DOE
by LLNL under contract W-7405-Eng-48.

\end{document}